\documentclass[review]{elsarticle}

\usepackage{lineno,hyperref}
\modulolinenumbers[5]
\usepackage{booktabs}
\usepackage{algorithm}
\usepackage{graphics,amsfonts,flushend}
\usepackage{times}
\usepackage{subfigure,epsfig,amsmath,amssymb}
\usepackage{setspace}
\usepackage{comment}
\usepackage{arydshln}
\usepackage{ragged2e}
\usepackage{threeparttable}
\usepackage{multicol}
\usepackage{algpseudocode}
\usepackage{color}
\usepackage[utf8]{inputenc}
\usepackage{listings}
\usepackage{xcolor}
\DeclareMathOperator{\tr}{tr}
\DeclareMathOperator{\diag}{diag}

\journal{Journal of The Franklin Institute}

\begin{document}

\begin{frontmatter}

\title{Interpretable Fault Detection using Projections of Mutual Information Matrix}
%\tnotetext[mytitlenote]{Fully documented templates are available in the elsarticle package on \href{http://www.ctan.org/tex-archive/macros/latex/contrib/elsarticle}{CTAN}.}

%% Group authors per affiliation:
%\author{Feiya Lv, Shujian Yu, Chenglin Wen, Jose C. Principe}
%\address{Radarweg 29, Amsterdam}
%\fntext[myfootnote]{Since 1880.}

%% or include affiliations in footnotes:
\author[mymainaddress]{Feiya Lv}
\ead{lvfeiya0215@126.com}

\author[mysecondaryaddress]{Shujian Yu\texorpdfstring{\corref{mycorrespondingauthor}}{Lg}}
\cortext[mycorrespondingauthor]{Corresponding author.}
\ead{yusjlcy9011@ufl.edu}

\author[mythirdaryaddress]{Chenglin Wen}
\ead{wencl@hdu.edu.cn}

\author[mysecondaryaddress]{Jose C. Principe}
\ead{principe@cnel.ufl.edu}

\address[mymainaddress]{School of Software Engineering, Anyang Normal University, Anyang 455000, PR China}
\address[mysecondaryaddress]{Department of Electrical and Computer Engineering, University of Florida, Gainesville, FL 32611, USA}
\address[mythirdaryaddress]{School of Automation, Hangzhou Dianzi University, Hangzhou 310018, PR China}

\begin{abstract}
This paper presents a novel mutual information (MI) matrix based method for fault detection. Given a $m$-dimensional fault process, the MI matrix is a $m \times m$ matrix in which the $(i,j)$-th entry measures the MI values between the $i$-th dimension and the $j$-th dimension variables. We introduce the recently proposed matrix-based R{\'e}nyi's $\alpha$-entropy functional to estimate MI values in each entry of the MI matrix. The new estimator avoids density estimation and it operates on the eigenspectrum of a (normalized) symmetric positive definite (SPD) matrix, which makes it well suited for industrial process.
We combine different orders of statistics of the transformed components (TCs) extracted from the MI matrix to constitute the detection index, and derive a simple similarity index to monitor the changes of characteristics of the underlying process in consecutive windows. We term the overall methodology ``projections of mutual information matrix" (PMIM).
Experiments on both synthetic data and the benchmark Tennessee Eastman process demonstrate the interpretability of PMIM in identifying the root variables that cause the faults, and its superiority in detecting the occurrence of faults in terms of the improved fault detection rate (FDR) and the lowest false alarm rate (FAR). The advantages of PMIM is also less sensitive to hyper-parameters. The advantages of PMIM is also less sensitive to hyper-parameters. Code of PMIM is available at \url{https://github.com/SJYuCNEL/Fault_detection_PMIM}.

\end{abstract}


\end{frontmatter}

%\linenumbers

\section{Introduction}

With the growing demand for security equipments and high-quality products, process monitoring has received tremendous attention in both academia and industry in the past decades. Fault detection, defined as the identification of abnormal operating conditions in real time, is an active topic in process monitoring. Data driven approaches have been the main stream for fault detection and control in recent years because they don't require neither a model not a priori information~\cite{Yin1S,Yin2S}. The multivariate statistical process monitoring (MSPM) is a well-known data-driven approach, and has been widely used in complex industrial environments~\cite{Macgregor1995J,Mason2002R,Wang2018Y}.

Traditional MSPM methods, e.g., principal component analysis (PCA)~\cite{Wise1990B}, partial least squares (PLS)~\cite{Kresta_1991} and independent component analysis (ICA)~\cite{LeeJM2004}, take advantage of the Hotteling $T^2$ statistic in principal component subspace or the squared prediction error (SPE) statistic in residual subspace to monitor the sample stream~\cite{Qin_2003,Hotelling_1936}. Although this kind of methods perform satisfactorily in the case of highly correlated multi-modal variables, they always neglect the temporal correlation between consecutive samples. Consequently, they cause a large Type-II error (i.e., fails to reject a false null-hypothesis).

To circumvent this limitation, the dynamic PCA (DPCA)~\cite{Ku_1995,YNDong2018}, the modified ICA (MICA)~\cite{CTong_2017,YWZhang2010,TongC2017} and various other recursive MSPM methods (e.g.,~\cite{Alcala_2009,LZhang_2016,ZWChen2017,Qingchao2017}) have been proposed thereafter. These methods usually add time-lagged variables in a sliding window to form a data matrix that captures the (local) dynamic characteristics of the underlying process. Compared with the traditional PCA or ICA, window-based methods distinguish better sample measurement from noise, thus offering a reliable avenue to address challenges associated with continuous processes~\cite{Wang_2010,SMZhang_2019}.

To further improve the performance of the above window-based methods, efficient extraction of high-order statistics of process variables is crutial~\cite{Choudhury_2004,Wang_2011,Shang_2017,Shang_2018,SMZhang_2019,BQZhou_2020}. Notable examples include statistics pattern analysis (SPA)~\cite{Wang_2010, Wang_2011}, recursive transformed component statistical analysis (RTCSA)~\cite{Shang_2017} and recursive dynamic transformed component statistical analysis (RDTCSA)~\cite{Shang_2018}. Different from traditional PCA and DPCA that implicitly assume that the latent variables follow a multivariate Gaussian distribution, SPA integrates the skewness, the kurtosis, and various other high-order statistics of the process measurement in sliding windows to deal with non-Gaussian data, demonstrating superior performance over PCA and DPCA. However, SPA performs poorly in case of incipient faults~\cite{Shang_2017}. To address this limitation, RTCSA and RDTCSA avoid dividing the projected space into principal component subspace and residual subspace. Instead, both methodologies take advantage of the full space to extract orthogonal transformed components (TCs), and evaluate a test statistic by incorporating the mean, the variance, the skewness, and the kurtosis of TCs. One should note that, the third- and forth-order information is usually beneficial to detect incipient faults \cite{Choudhury_2004,Wang_2010,Wang_2011,Shang_2017,Shang_2018,SMZhang_2019}. Although RTCSA and RDTCSA enjoy solid mathematical foundation, the TCs from a covariance matrix only capture linear relationships among different dimensions of measurement. Therefore, a reliable way to extract nonlinear statistics among different dimensions of measurements becomes a pivotal problem in fault detection~\cite{Jia_2016,Lv_2018,Chang_2015}.

The application of information theory on fault detection is an emerging and promising topic~\cite{Bazan_2017,XiaoZhao_2017}. Although there are a few early efforts that attempt to shed light on fault detection with information-theoretic concepts, they simply employ (an approximation to) the MI to select a subset of the most informative variables to circumvent the curse of dimensionality (e.g.,~\cite{Verron_2008,Jiang_2011,MMRashid2012,Yu_2013,Jiang_2018,Joshi_2005}). To the best of our knowledge, there are only two exceptions that illuminate the potential of using information-theoretic concepts for fault detection, beyond the role of variable selection. Unfortunately, no specific statistical analysis is presented~\cite{Jiang_2018,Joshi_2005}. Therefore, the design from first principles of a fault detection method using information theory remains an open problem\footnote{Note that, this work does not use the physical significance of entropy, which was initially introduced in thermodynamics. According to Boltzmann the function of entropy can be expressed as: $S=-k\ln p$, where $k$ is Boltzmann constant, $p$ is thermodynamic probability. Instead, this work is based on information entropy by Shannon in 1948~\cite{shannon1948mathematical}, which was used to measure the uncertainty of signal source in a transmission system.}. The detailed contribution of this work is multi-fold:
\begin{itemize}
	\item \textbf{Novel methodology:} We construct a MI matrix to monitor the (possibly nonlinear) dynamics and the non-stationarity of the fault process. A novel fault detection method, i.e., projections of mutual information matrix (PMIM), is also developed thereafter.
	\item \textbf{Novel estimator:} Unlike previous information-theoretic fault detection methods which usually use the classical Shannon entropy functional that relies heavily on the precise estimation of underlying data distributions, we suggest using the recently proposed matrix-based R{\'e}nyi's $\alpha$-entropy functional to estimate MI values. The new estimator avoids estimation of the underlying probability density function (PDF), and employs the eigenspectrum of a (normalized) symmetric positive definite (SPD) matrix. This intriguing property makes the novel estimator easily applicable to real-world complex industrial process which usually contains continuous, discrete and even mixed variables.
    \item \textbf{Detection accuracy:} Experiments on both synthetic data and the benchmark Tennessee Eastman process (TEP) indicate that PMIM achieves comparable or slightly higher detection rates than state-of-the-art fault detection methods. Moreover, PMIM enjoys significantly lower false detection rate.
    \item \textbf{Implementation details and reproducibility:} We elaborate the implementation details of fault detection using PMIM. We also illustrate the detectability of PMIM using the eigenspectrum of the MI matrix. For reproducible results, we provide key functions (in MATLAB $2019$a) concerning PMIM in the Appendix A. We also release a full demo of PMIM at \url{https://github.com/SJYuCNEL/Fault_detection_PMIM}.
    \item \textbf{Interpretability:} Fault detection using PMIM can provide insights on the the exact root variables that lead to the occurrence of fault. In this sense, the result of fault detection using PMIM is interpretable, i.e., the practitioners know which variable or specific sensor data causes the fault.
\end{itemize}

The remainder of this paper is organized as follows. We first introduce the definition of MI matrix and present its estimation with the matrix-based R{\'e}nyi's entropy functional in Section 2. We then describe our proposed fault detection using PMIM in Section 3, and elaborate its implementation details in Section 4. Experiments on both synthetic and TEP benchmark are performed in Section 5. We finally conclude this work and discuss future directions in Section 6.

\emph{Notations:} Throughout this paper, scalars are denoted by lowercase letters (e.g., $x$), vectors appear as lowercase boldface letters (e.g., $\mathbf{x}$), and matrices are indicated by uppercase letters (e.g., $X$). The $(i,j)$-th element of $X$ is represented by $X_{ij}$. If $X$ is a square matrix, then $X^{-1}$ denotes its inverse. $I$ stands for the identity matrix with compatible dimensions. The $i$-th row of a matrix $X$ is declared by the row vector $\mathbf{x}^i$, while the $j$-th column is indicated with the column vector $\mathbf{x}_j$. Moreover, superscript indicates time (or sample) index, subscript indicates variable index. For $\mathbf{x}\in \mathbb{R}^n$, the $\ell_p$-norm of $\mathbf{x}$ is defined as $\|\mathbf{x}\|_p\triangleq(\sum\limits_{i=1}^n|x_i|^p)^{\frac{1}{p}}$.

\section{The MI Matrix: Definition and Estimation}

\subsection{The Definition of MI matrix}
MI quantifies the nonlinear dependence between two random variables~\cite{Cover1991,Latham2009}. Therefore, given a multivariate time series (here refers to fault process), an MI matrix (in a stationary environment) can be constructed by evaluating MI values between each pair of variables. Intuitively, the MI matrix can be viewed as a nonlinear extension of the classical covariance matrix. Specifically, the formal definition of MI matrix is given as follows.

% Given a $m$ dimensional process that is characterized by $n$ samples, $\wp=\{\mathbf{\iota}_{1}, \mathbf{\iota}_{2}, \cdots, \mathbf{\iota}_{m}\}\in\mathbb{R}^{n \times m}$, $\mathbf{\iota}_{i}\in\mathbb{R}^{n}$, $i=1,2,\cdots,n$. Suppose $\wp$ is stationary

\noindent\textbf{Definition 1.} Given a $m$-dimensional (stationary) process $\wp$, let us denote $\mathbf{x}_i$ ($i=1,2,\cdots,m$) the $i$-th dimensional of the process measurement, then the MI matrix over $\wp$ is defined as:
\begin{equation}\label{eq_MIdefine}
M=\begin{bmatrix}
I(\mathbf{x}_{1}; \mathbf{x}_{1}) & I(\mathbf{x}_{1}; \mathbf{x}_{2}) & \cdots & I(\mathbf{x}_{1}; \mathbf{x}_{m})\\
I(\mathbf{x}_{2}; \mathbf{x}_{1}) & I(\mathbf{x}_{2}; \mathbf{x}_{2}) & \cdots & I(\mathbf{x}_{2}; \mathbf{x}_{m})\\
\vdots & \vdots & \ddots & \vdots \\
I(\mathbf{x}_{m}; \mathbf{x}_{1}) & I(\mathbf{x}_{m}; \mathbf{x}_{2}) & \cdots & I(\mathbf{x}_{m}; \mathbf{x}_{m})\\
\end{bmatrix} \in \mathbb{R}^{m \times m},
\end{equation}
where $I(\mathbf{x}_{i}; \mathbf{x}_{j})$ denotes MI between variables $\mathbf{x}_{i}$ and $\mathbf{x}_{j}$.

According to Shannon information theory~\cite{shannon1948mathematical}, $I(\mathbf{x}_i;\mathbf{x}_j)$ is defined over the joint probability distribution of $\mathbf{x}_i$ and $\mathbf{x}_j$ (i.e., $p(\mathbf{x}_i,\mathbf{x}_j)$) and their respectively marginal distributions (i.e., $p(\mathbf{x}_i)$ and $p(\mathbf{x}_j)$). Specifically,
\begin{equation}
\label{eq_Shannon}
\begin{split}
I(\mathbf{x}_{i}; \mathbf{x}_{j}) &\!=\!\!\int\!\!\int\!\! p(\mathbf{x}_i,\mathbf{x}_j)\log\left(\frac{p(\mathbf{x}_i,\mathbf{x}_j)}{p(\mathbf{x}_i)p(\mathbf{x}_j)}\right)d\mathbf{x}_i d\mathbf{x}_j\\
& = - \int\!\!\left(\!\!\int\!\!p(\mathbf{x}_i,\mathbf{x}_j)d\mathbf{x}_j\!\!\right)\!\!\log p(\mathbf{x}_i)d\mathbf{x}_i\!\!-\!\!\int\!\!\left(\!\!\int\!\! p(\mathbf{x}_i,\mathbf{x}_j)d\mathbf{x}_i\!\!\right)\!\!\log p(\mathbf{x}_j)d\mathbf{x}_j \\
& ~~~~ + \int\int p(\mathbf{x}_i,\mathbf{x}_j)\log p(\mathbf{x}_i,\mathbf{x}_j)d\mathbf{x}_i d\mathbf{x}_j\\
& = - \int\!\!p(\mathbf{x}_i)\log p(\mathbf{x}_i)d\mathbf{x}_i\!-\!\!\int\!\!p(\mathbf{x}_j)\log p(\mathbf{x}_j)d\mathbf{x}_j\!+\!\!\int\!\!\int\!\! p(\mathbf{x}_i,\mathbf{x}_j)\log p(\mathbf{x}_i,\mathbf{x}_j)d\mathbf{x}_i d\mathbf{x}_j\\
&\!\!=\!\!H(\mathbf{x}_{i})\!+\!H(\mathbf{x}_{j})\!-\!H(\mathbf{x}_{i}, \mathbf{x}_{j}),
\end{split}
\end{equation}
where $H(\cdot)$ denote the entropy and $H(\cdot, \cdot)$ denotes the joint entropy. In particular, $I(\mathbf{x}_{i}; \mathbf{x}_{i})=H(\mathbf{x}_{i})$.

Theoretically, the MI matrix is symmetric and non-negative\footnote{By applying the Jensen inequality, we have \\ $I(\mathbf{x}_i; \mathbf{x}_j)\!\!=\!\!\int\!\!\!\int p(\mathbf{x}_i, \mathbf{x}_j)\log\left(\frac{p(\mathbf{x}_i, \mathbf{x}_j)}{p(\mathbf{x}_i)p(\mathbf{x}_j)}\right)d\mathbf{x}_i d\mathbf{x}_j \geq \!\!-\!\!\log\left(\int\!\!\!\int p(\mathbf{x}_i,\mathbf{x}_j)\left(\frac{p(\mathbf{x}_i)p(\mathbf{x}_j)}{p(\mathbf{x}_i,\mathbf{x}_j)}\right)\right) \!\!=\!\! -\log(\int\!\!\!\int p(\mathbf{x}_i)p(\mathbf{x}_j))=0$.}. Moreover, in the absence of any dependence in pairwise variables, the MI matrix reduces to a diagonal matrix with the entropy of each variable lies on the main diagonal. Interestingly, although the estimated MI matrix has been conjectured and also observed in our application to be positive semidefinite, this property is not always true theoretically~\cite{Jakobsen_2014}.

\subsection{Estimate MI matrix with matrix-based R{\'e}nyi's \texorpdfstring{$\alpha$}{Lg}-order entropy}

Entropy measures the uncertainty in a random variable using a single scalar quantity~\cite{Principe_2010,Cover_2017}. For a random variable (or vector) $\mathbf{x}$, with probability density function (PDF) $p(\mathbf{x})$ in a finite set $\mathbf{s}$, a natural extension of the Shannon's differential entropy is the R{\'e}nyi's $\alpha$-order entropy~\cite{Lennert_2013}:
\begin{equation}\label{eq_renyi}
H_{\alpha}(\mathbf{x}) =\frac{1}{1-\alpha}\log \int_{\mathbf{s}} p^{\alpha}(\mathbf{x})d\mathbf{x}.
\end{equation}
It is well-known that, when $\alpha\to 1$, Eq.~(\ref{eq_renyi}) reduces to the basic Shannon's differential entropy\footnote{A simple proof by applying the L'H\^{o}spital's rule at $\alpha=1$ is shown in~\cite{bromiley2004shannon}.} $H(\mathbf{x})=-\int _{\mathbf{s}} p(\mathbf{x}) \log p(\mathbf{x})d\mathbf{x}$. In this perspective, R{\'e}nyi's entropy makes a one-parameter generalization to the basic Shannon definition by introducing a hyperparameter $\alpha$.

Information theory has been successfully applied to various machine learning, computer vision and signal processing tasks~\cite{Principe_2010,yu2019multivariate}. Unfortunately, the accurate PDF estimation in Eq.~(\ref{eq_renyi}) on continuous and complex data impedes its more widespread adoption in data driven science. This problem becomes more severe for process monitoring, since the obtained multivariate measurement may contain both discrete and continuous variables. Moreover, there is still no universal agreement on the definition of MI between discrete and continuous variables~\cite{ross2014mutual,gao2017estimating}, let alone its precise estimation.
In this work, we use a novel estimator developed by S{\'a}nchez Giraldo \emph{et~al}.~\cite{Sanchez_2015} to estimate the MI matrix. Specifically, according to~\cite{yu2019multivariate,Sanchez_2015}, it is feasible to evaluate a quantity that resembles quantum R{\'e}nyi's entropy~\cite{Lennert_2013} in terms of the normalized eigenspectrum of the Hermitian matrix of the projected data in reproducing kernel Hilbert space (RKHS), thus estimating the entropy directly from data without PDF estimation. For completeness, we provide below S{\'a}nchez Giraldo \emph{et al}.'s definition on entropy and joint entropy.

\noindent\textbf{Definition 2.} Let $\kappa:\chi \times \chi \mapsto \mathbb{R}$ be a real valued positive definite kernel that is also infinitely divisible~\cite{Bhatia_2006}. Given $\{\mathbf{x}_{i}\}_{i=1}^{n}\in \chi$, each $\mathbf{x}_i$ can be a real-valued scalar or vector, and the Gram matrix  $K$ obtained from evaluating a positive definite kernel $\kappa$ on all pairs of exemplars, that is $K=\kappa(\mathbf{x}_{i}, \mathbf{x}_{j})$, a matrix-based analogue to R{\'e}nyi's $\alpha$-entropy for a normalized positive definite matrix $A$ of size $n\times n$, such that $\tr(A)=1$, can be given by the following functional:
\begin{equation}\label{def_entropy}
H_{\alpha}(A)=\frac{1}{1-\alpha}\log \left(\tr(A^{\alpha})\right)=\frac{1}{1-\alpha}\log_{2}\left(\sum_{i=1}^{n}\lambda _{i}(A)^{\alpha}\right),
\end{equation}
where $A$ is the normalized version of $K$, i.e., $A=K/{\text{tr}(K)}$,  and $\lambda _{i}(A)$ denotes the $i$-th eigenvalue of $A$.

\noindent\textbf{Definition 3.} Given $n$ pairs of samples $(\mathbf{x}_{i}, \mathbf{y}_{i})_{i=1}^{n}$, each sample contains two different types of measurements $\mathbf{x}\in\chi$ and $\mathbf{y}\in\gamma$ obtained from the same realization, and the positive definite kernels $\kappa_{1}:\chi\times\chi\mapsto\mathbb{R}$ and $\kappa_{2}:\gamma\times\gamma\mapsto\mathbb{R}$ , a matrix-based analogue to R{\'e}nyi's $\alpha$-order joint-entropy can be defined as:
\begin{equation}\label{def_joint_entropy}
H_{\alpha}(A,B)=H_{\alpha}\left(\frac{A \circ B}{\tr(A \circ B)}\right),
\end{equation}
where $A_{ij}=\kappa_{1}(\mathbf{x}_{i}, \mathbf{x}_{j})$ , $B_{ij}=\kappa_{2}(\mathbf{y}_{i}, \mathbf{y}_{j})$ and $A\circ B$  denotes the Hadamard product between the matrices $A$ and $B$.

Given Eqs.~(\ref{def_entropy})-(\ref{def_joint_entropy}), the matrix-based R{\'e}nyi's $\alpha$-order MI $I_{\alpha}(A; B)$ in analogy of Shannon's MI is given by:
\begin{equation}\label{def_mutual}
I_{\alpha}(A;B)=H_{\alpha}(A)+H_{\alpha}(B)-H_{\alpha}(A,B).
\end{equation}
Throughout this paper, we use the Gaussian kernel $\kappa(\mathbf{x}_{i},\mathbf{x}_{j})=\exp(-\frac{\|\mathbf{x}_{i}-\mathbf{x}_{j}\|^{2}}{2\sigma ^{2}})$ to obtain the Gram matrices. Obviously, Eq.~(\ref{def_mutual}) avoids real-valued PDF estimation and has no additional requirement on data characteristics (e.g., continuous, discrete, or mixed), which makes it has great potential in our application.

\section{The Fault Detection using PMIM}

\begin{figure}[!t]
  \centering
  \includegraphics[width=0.9\textwidth]{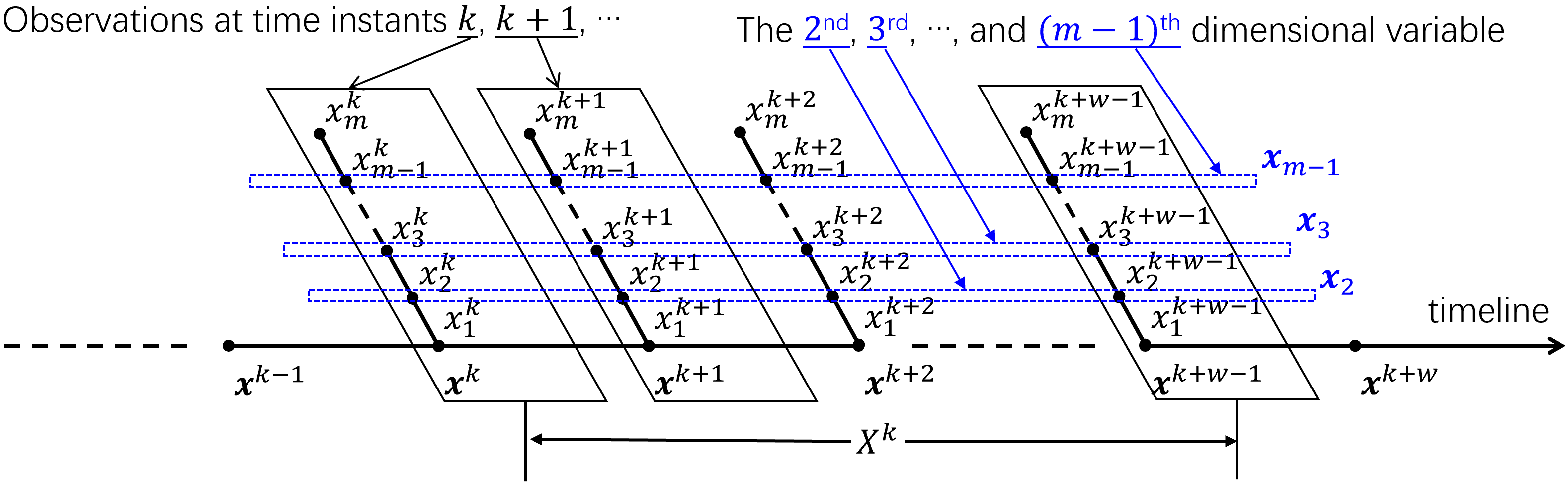}
  \caption{Local sample matrix with a sliding window of size $w$.}\label{fig:windowX}
\end{figure}

In this section, we present PMIM, a novel fault detection method by monitoring the statistics associated with the MI matrix. Given a discrete time process $\aleph=\{\mathbf{x}^1,\mathbf{x}^2,\cdots\}:\mathbf{x}^{i} \in \mathbb{R}^{1 \times m}$, at each time instant $k$, we construct a local sample matrix $X^k\in \mathbb{R}^{w\times m}$ of the following form:
\begin{equation}\label{eq_sample_matrix}
\begin{split}
X^k &=\begin{bmatrix}
\mathbf{x}^{k-w+1}\\
\mathbf{x}^{k-w+2}\\
\vdots \\
\mathbf{x}^{k}
\end{bmatrix}
=\begin{bmatrix}
x^{k-w+1}_1& x^{k-w+1}_2& \cdots& x^{k-w+1}_m\\
x^{k-w+2}_1& x^{k-w+2}_2& \cdots& x^{k-w+2}_m\\
\vdots& \vdots& \ddots& \vdots\\
x^{k}_1& x^{k}_2& \cdots& x^{k}_m
\end{bmatrix}\\
&\triangleq
\left[
    \begin{array}{c;{2pt/2pt}c;{2pt/2pt}c;{2pt/2pt}c}
        \mathbf{x}_1 & \mathbf{x}_2 &\cdots  & \mathbf{x}_m
    \end{array}
\right]
 \in \mathbb{R}^{w \times m},
\end{split}
\end{equation}
where $\mathbf{x}_j$ ($1\leq j\leq m$) denotes the $j$-th dimensional variable that is characterized by $w$ realizations. Fig.~\ref{fig:windowX} illustrates $\mathbf{x}^i$, $\mathbf{x}_j$ and $X$. Each variable is mean centered and normalized to $[0,1]$ to account for different value ranges~\cite{Macgregor1995J,Mason2002R,Wang2018Y,Wise1990B,Kresta_1991}. Then the MI matrix $M$ at time instant $k$ is given by:
\begin{equation} \label{eq_MI_matrix}
M=\begin{bmatrix}
H(\mathbf{x}_{1}) & I(\mathbf{x}_{1}; \mathbf{x}_{2}) & \cdots & I(\mathbf{x}_{1}; \mathbf{x}_{m})\\
I(\mathbf{x}_{2}; \mathbf{x}_{1}) & H(\mathbf{x}_{2}) & \cdots & I(\mathbf{x}_{2}; \mathbf{x}_{m})\\
\vdots & \vdots & \ddots & \vdots \\
I(\mathbf{x}_{m}; \mathbf{x}_{1}) & I(\mathbf{x}_{m}; \mathbf{x}_{2}) & \cdots & H(\mathbf{x}_{m})
\end{bmatrix} \in \mathbb{R}^{m \times m}\\.
\end{equation}

The general idea of our method is that $M$ contains all the nonlinear dependencies between any pairwise variables of the underlying fault process at time instant $k$. In a stationary environment, any quantities or statistics associated with $M$ should remain unchanged or stable. However, the existence of an abrupt fault may affect, at least, the values of one or more entries in the MI matrix, thus altering the values of our monitored quantities or statistics extracted from MI matrix.

Prior art suggests that those reliable quantities can be extracted from the orthogonal space spanned by eigenvectors of the sample covariance matrix (e.g.,~\cite{Wise1990B,Kresta_1991,Ku_1995,Lee_2006,Shang_2017,Shang_2018,BQZhou_2020}).
Motivated by this idea, suppose the eigenvalue decomposition of MI matrix is given by $M=P\Lambda P^{-1}$, where $P\in\mathbb{R}^{m\times m}$  is the matrix of eigenvectors and $\Lambda = \diag(\lambda_1,\lambda_2,\cdots,\lambda_m)\in \mathbb{R}^{m \times m}$ is a diagonal matrix with eigenvalues on the main diagonal. Then, a new representation of $X$ (denote it $T$) in the orthogonal space spanned by column vectors in $P$ can be expressed as,
\begin{equation} \label{eq_TC}
T= X P\triangleq
\begin{bmatrix}
        \mathbf{t}^{k-w+1} \\ \mathbf{t}^{k-w+2} \\ \vdots  \\ \mathbf{t}^k
\end{bmatrix}
\in \mathbb{R}^{w \times m}.
\end{equation}

We term the column vectors of $T$ the mutual information based transform components (MI-TCs). The terminology of transform components (TCs) originates from~\cite{Wise1990B,Kresta_1991,Shang_2017} and is defined over the sample covariance matrix $C=\frac{1}{w-1}X^{T}X$. Specifically, suppose $P_{C}$ and $\Lambda_{C}$ are respectively the eigenvectors and eigenvalues of $C$, i.e., $C=P_{C}\Lambda_{C} {P_{C}}^{-1}$, then the original TCs of $X$ are given by $T_{C}=XP_{C}\in \mathbb{R}^{w\times m}$.

\begin{figure}[!ht]
\centering
\includegraphics[width=0.95\textwidth]{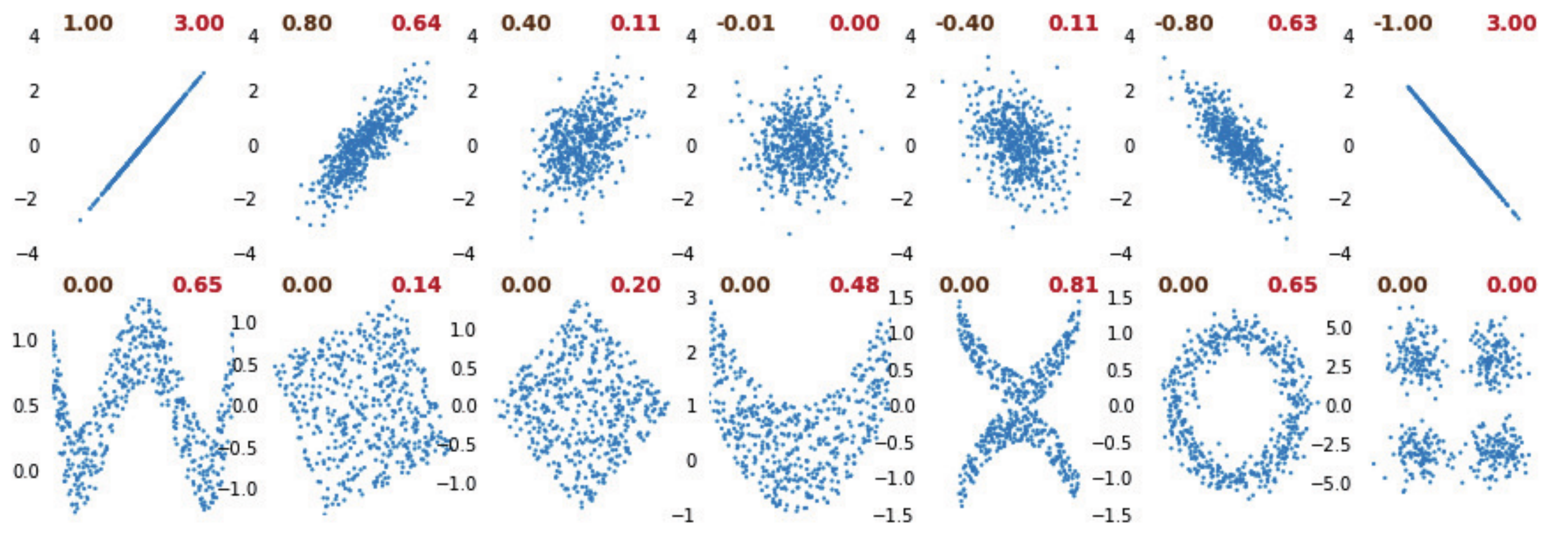}\\
\caption{Examples of correlation versus mutual information (MI) estimated by the classic Shannon's discrete entropy functional with the formula $H(\mathbf{x})=-\sum \limits_{x\in \mathbf{x}}p(x)\log_{2}p(x)$, over 500 samples. Each panel illustrates a scatter plot of samples drawn from a particular bivariate distribution. For each example, the correlation between the two variables is shown in brown (left) and the MI is shown in red (right). The top row shows linear relationships, for which MI and correlation both detect a relationship (although in different scales). The bottom row shows a series of distributions for which the correlation is $0$, but the MI is significant larger than $0$.}\label{fig:corMI}
\end{figure}

Compared with the MI matrix $M$, the covariance matrix $C$ only captures the linear dependence (correlation) between pairwise dimensions of the normalized measurement~\cite{Shang_2017}. By contrast, the MI matrix $M$ operates with the full PDF information between pairs of variables and makes no assumption on the joint distribution of the measurement nor the nature of the relationship
between pairwise dimensions. Moreover, it can simply identify nonlinear and non-monotonic dependencies \cite{InceRAZ2017},
which are common in industrial process~\cite{Hotelling_1936,Shang_2017,BQZhou_2020,Dyson_1962}. See Fig.~\ref{fig:corMI} for a few concrete examples on the advantage of MI over linear correlation, in which the linear correlation fails completely in quantifying nonlinear and non-monotonic effects (the bottom row).

%By contrast, the MI matrix $M$ is able to detect highly nonlinear and non-monotonic effects between pairwise dimensions~\cite{InceRAZ2017}.

%\textcolor{blue}{Each panel illustrates a scatter plot of samples drawn from a particular bivariate distribution. For each example, the correlation between the two variables is shown in brown and the MI is shown in red. For linear relationships (top row), the MI and correlation both detected the relationship. However, there are some distributions of correlation zero (bottom row), their relationships can still be detected by MI.}

In each sliding window, we characterize $T$ with a detection index $\mathbf{\Theta}^{k}=[\mathbf{\mu}_{k}|\mathbf{\nu}_{k}|\mathbf{\zeta}_{k}|\mathbf{\gamma}_{k}]^{T}\in \mathbb{R}^{4m}$, it consists of the first-order statistic (i.e., the mean $\mathbf{\mu}_{k}=\mathbb{E}(\mathbf{t}^{k})$), the second-order statistic (i.e., the variance $\mathbf{\nu}_{k}=\mathbf{\sigma}_{k}^{2}=\mathbb{E}\left[({\mathbf{t}^{k}-\mathbf{\mu}_{k}})^2\right]$), the third-order statistic (i.e., the skewness $\mathbf{\zeta}_{k}=\mathbb{E}\left[\left(\frac{\mathbf{t}^{k}-\mathbf{\mu}_{k}}{\mathbf{\sigma}_{k}}\right)^3\right]$), and the forth-order statistic (i.e., the excess kurtosis $\mathbf{\gamma}_{k}=\mathbb{E}\left[\left(\frac{\mathbf{t}^{k}-\mathbf{\mu}_{k}}{\mathbf{\sigma}_{k}}\right)^4\right]-3$).
Specifically, the empirical estimation to $\mathbf{\mu}_{k}$, $\mathbf{\nu}_{k}$, $\mathbf{\zeta}_{k}$ and $\mathbf{\gamma}_{k}$ are given by:
\begin{equation}\label{eq_mean}
\mathbf{\mu}_{k}= \frac{1}{w}\sum _{i=0}^{w-1}\mathbf{t}^{k-i}\in \mathbb{R}^{1 \times m},
\end{equation}
\begin{equation}\label{eq_variance}
\mathbf{\nu}_{k}=\frac{1}{w}\sum _{i=0}^{w-1}\left(\mathbf{t}^{k-i}-\mathbf{\mu}_{k}\right)^{2}\in \mathbb{R}^{1\times m},
\end{equation}
\begin{equation} \label{eq_skewness}
\mathbf{\zeta}_{k}=\frac{1}{w\mathbf{\sigma}_{k}^{3}}\sum _{i=0}^{w-1}\left(\mathbf{t}^{k-i}-\mathbf{\mu}_{k}\right)^{3}\in \mathbb{R}^{1 \times m},
\end{equation}
\begin{equation}\label{eq_kurtosis}
\mathbf{\gamma}_{k}=\frac{1}{w\mathbf{\sigma}_{k}^{4}}\sum _{i=0}^{w-1}\left(\mathbf{t}^{k-i}-\mathbf{\mu}_{k}\right)^{4}-3\in \mathbb{R}^{1 \times m}.
\end{equation}

Note that, $\mathbf{\mu}^{*}= \mathbb{E}\left[\mu_{k}\right]$ (the mean of the TCs under normal condition) is used for the online calculation of detection index. When a fault occurs, one or more of the four statistics (namely, $\mathbf{\mu}_k, \mathbf{\nu}_k, \mathbf{\zeta}_k$ and $\mathbf{\gamma}_k$) are expected to deviate significantly from their expectations.

Given $\Theta^k$, a similarity index for local sample matrix $X^k$ at time instant $k$ can be defined as:
\begin{equation}\label{similarity}
D^k=\|\Theta_{\sigma}^{-1}(\Theta^{k}-\Theta_{\mu})\|_{p},
\end{equation}
where $\Theta_{\mu}$ denotes the mean value of similarity index over training data, $\Theta_\sigma=\diag(\sigma_1,\sigma_2,\cdots,\sigma_{4m})$ denotes a diagonal matrix in which the main diagonal consists of the standard deviation in each dimension of $\Theta^k$. The empirical method based on training data is used to determine the upper control limit $D_{\text{cl}}$ with a given confidence level $\eta$~\cite{Wang_2010}. An online monitoring procedure is then used to quantify the dissimilarity of statistics between normal and abnormal states.

Algorithm 1 and Algorithm 2 summarize, respectively, the offline training and the online testing of our proposed PMIM.
\begin{algorithm} [!ht]
\caption{Fault detection using PMIM (training phase)}
\small
\label{PermutationAlg1}
\begin{algorithmic}[1]
\Require
Process measurements $\aleph=\{\mathbf{x}^{i}|\mathbf{x}^{i} \in \mathbb{R}^{m} \}_{i=1}^{n} $;
sliding window size $w$;
significance level $\eta$.
\Ensure
mean of the transform components (TCs) $\mathbf{\mu}^{*}$;
standard deviation $\Theta_{\sigma}$ of the detection index;
reference mean $\Theta_{\mu}$ of the detection index.
\For {$i = 1$ to $ n$}
\State Construct a local time-lagged matrix $X^{i}\in \mathbb{R}^{w\times m}$ at time instant $i$ by Eq.~(\ref{eq_sample_matrix});
\State Construct the MI matrix $M^{i}$ by Eq.~(\ref{eq_MI_matrix});
%\State Perform eigenvalue decomposition to $M^i=P\Lambda P^{-1}$;
\State Obtain the TCs $T^{i}$ of $X^{i}$ by Eq.~(\ref{eq_TC});
\State Obtain the detection index $\Theta^{i}=[\mathbf{\mu}_{i}|\mathbf{\nu}_{i}|\mathbf{\zeta}_{i}|\mathbf{\gamma}_{i}]^{T}$ by Eqs.~(\ref{eq_mean})-(\ref{eq_kurtosis}).
\EndFor
\State Calculate the mean of the TCs $\mathbf{\mu}^{*}=\sum\limits_{i=1}^{n}{\mathbf{\mu}_{i}}$, reference mean $\Theta_{\mu}$ and standard deviation $\Theta_{\sigma}$.
\For {$i = 1$ to $ n$}
\State ${\kern 8pt}$ $D^i=\|\Theta_{\sigma}^{-1}(\Theta^{i}-\Theta_{\mu})\|_{p}$.
\EndFor
\State Determine the control limit $D_{\text{cl}}$ at the significance level $\eta$. \\
\Return $\mathbf{\mu}^{*}$; $\Theta_{\sigma}$; $\Theta_{\mu}$; $D_{\text{cl}}$
\end{algorithmic}
\end{algorithm}

\begin{algorithm} [!ht]
\caption{Fault detection using PMIM (testing phase)}
\small
\label{PermutationAlg2}
\begin{algorithmic}[1]
\Require
The online process measurement $\{\mathbf{x}_{\text{test}}^1,\mathbf{x}_{\text{test}}^2,\cdots\}$; sliding window size $w$;
mean of the transform components (TCs) $\mu^{*}$;
standard deviation $\Theta_{\sigma}$ of the detection index;
reference mean $\Theta_{\mu}$of the detection index;
control limit $D_{\text{cl}}$.
\Ensure
\emph{Decision:} alarm or not.
\While {\text{End of process not reached}}
\State Construct a local time-lagged matrix $X_{\text{test}}^{i}\in \mathbb{R}^{w\times m}$ at time instant $i$ by Eq.~(\ref{eq_sample_matrix});
\State Construct the MI matrix $M_{\text{test}}^{i}$ by Eq.~(\ref{eq_MI_matrix});
\State Obtain the TCs $T_{\text{test}}^{i}$ of $ X_{\text{test}}^{i}$ by Eq.~(\ref{eq_TC});
\State Obtain the detection index $\Theta_{\text{test}}^{i}=[\mathbf{\mu}_{i}|\mathbf{\nu}_{i}|\mathbf{\zeta}_{i}|\mathbf{\gamma}_{i}]_{\text{test}}^{T}$ with the mean of the TCs $\mathbf{\mu}^{*}$;
\State Obtain the similarity index by $D_{\text{test}}^{i}=\|\Theta_{\sigma}^{-1}(\Theta_{\text{test}}^{i}-\Theta_{\mu})\|_{p}$;
\If {$D_{\text{test}}^{i} \geq D_{\text{cl}}$}
\State Alarm the occurrence of fault;
\State Identify the root variables that cause the fault;
\Else
\State $i=i+1$; Go back to Step 2.
\EndIf
\EndWhile \\
\Return \emph{Decision}
\end{algorithmic}
\end{algorithm}

\section{A Deeper Insight into the Implementation of PMIM}

In this section, we elaborate the implementation details of PMIM. The discussion is based on a synthetic process with time-correlated dynamics\cite{Shang_2017,Shang_2018}:
\begin{equation}
\mathbf{x}= A\mathbf{s}+\mathbf{e},
\end{equation}
where $\mathbf{x}\in \mathbb{R}^{m}$ is the process measurements, $\mathbf{s}\in \mathbb{R}^{r}(r<m)$ is the data sources, $\mathbf{e}\in \mathbb{R}^{m}$ is the noise, and $A\in \mathbb{R}^{m\times r}$ is coefficient matrix that assumed to be column full rank \cite{Shang_2018,Alcala_2009}. Let us assume data sources satisfy the following relations:
\begin{equation}
s^k_i=\sum_{j=1}^{l}\beta_{i,j}v^{k-j+1}_i,
\end{equation}
where $s^k_i$ is the $i$-th variable at time $k$, $v^{k-j+1}_i$ represents the value of the $i$-th Gaussian data source with time independence at time $k-j+1$, $\beta_{i,j}$ denotes the weight coefficient, $l\geq 2$. Obviously, both $\mathbf{s}$ and $\mathbf{x}$ are time-correlated.

Here, the fault type of sensor bias\footnote{Other fault types, such as sensor precision degradation $\mathbf{x}^{*}=\eta \mathbf{x}$, gain degradation $\mathbf{x}^{*}=\mathbf{x}+\mathbf{\xi}_{m}\mathbf{e}^{[s]}$, additive process fault $\mathbf{x}=A(\mathbf{s}+\mathbf{\xi}_{m}\mathbf{f}^{[p]})+\mathbf{e}$ and dynamic changes $\tilde{\beta}=\beta+\triangle \beta$ can also analyzed similarly.} is considered:
\begin{equation}
\mathbf{x}^{*}=\mathbf{x}+\mathbf{f},
\end{equation}
where $\mathbf{x}^{*}$ is the measurement under sensor bias, and $\mathbf{x}$ denotes the fault-free portion. In the following, we will show how $\mathbf{f}$ affects the matrix-based R{\'e}nyi's $\alpha$-order entropy.

The matrix-based R{\'e}nyi's $\alpha$-order entropy is a non-parametric measure of entropy. For the $p$-th variable with $w$ realizations,
we build its Gram matrix $K\in \mathbb{R}^{w\times w}$ (at time instant $k$) by projecting it into a RKHS with an infinite divisible kernel\footnote{In this work, we simply use the radial basis function (RBF) kernel $G_{\sigma}(\cdot)=\exp(-\frac{\|\cdot\|^2}{2\sigma^2})$ as recommended in~\cite{Sanchez_2015,yu2019multivariate}.}:
\begin{equation}\label{K_normal}
\centering
K_{\mathbf{x}_p}=\begin{bmatrix}
   1 &\exp\!\left(\!-\frac{(x^{k-w+1}_{p}-x^{k-w+2}_{p})^{2}}{2\sigma ^{2}}\!\right)\! &\cdots &\exp\!\left(\!-\frac{(x^{k-w+1}_{p}-x^{k}_{p})^{2}}{2\sigma ^{2}}\!\right)\! \\
   \exp\!\left(\!-\frac{(x^{k-w+2}_{p}-x^{k-w+1}_{p})^{2}}{2\sigma ^{2}}\!\right)\! &1 &\cdots &\exp\!\!\left(\!\!-\frac{(x^{k-w+2}_{p}-x^{k}_{p})^{2}}{2\sigma ^{2}}\!\right)\! \\
   \vdots &\vdots &\ddots &\vdots \\
   \exp\!\left(\!-\frac{(x^{k}_{p}-x^{k-w+1}_{p})^{2}}{2\sigma ^{2}}\!\right)\! & \exp\!\left(\!-\frac{(x^{k}_{p}-x^{k-w+2}_{p})^{2}}{2\sigma ^{2}}\!\right)\! &\cdots &1\\
\end{bmatrix}.
\end{equation}

We normalize $K$ by its trace, i.e., $K=K/{\text{tr}(K)}$. It should be noted that the kernel induced mapping can be understood as a means of computation of high order statistics\footnote{By the Taylor expansion of the RBF kernel, we have \\ $\kappa({x}^{i},{x}^{j})=\exp\left(-\gamma\|{x}^{i}-{x}^{j}\|^{2}\right)=\exp\left(-\gamma{x^i}^{2}\right)\exp\left(-\gamma{x^j}^{2}\right)\left(1+\frac{2\gamma{x}^{i}{x}^{j}}{1!}+\frac{(2\gamma{x}^{i}{x}^{j})^2}{2!}+\frac{(3\gamma{x}^{i}{x}^{j})^2}{3!}+\cdots\right)$, where $\gamma=\frac{1}{2\sigma^2}$.}.

Suppose the fault occurs exactly at the $p$-th variable, i.e., $\mathbf{x}_{p}^{*}=\mathbf{x}_{p}+\mathbf{f}$ and $\mathbf{f}=\{f^{k-w+1}, f^{k-w+2}, \cdots, f^{k}\}$. The $(i,j)$-th entry of the Gram matrix $K$ associated with $\mathbf{x}_{p}$ becomes:
\begin{equation}
\begin{split}
\exp\!\left(\!-\frac{||x_{p}^{i*}-x_{p}^{j*}||^{2}}{2\sigma ^{2}}\!\right)\!
&=\exp\!\left(\!-\frac{[(x_{p}^{i}+f^{i})-(x_{p}^{j}+f^{j})]^{2}}{2\sigma ^{2}}\!\right)\!\\
&=\exp\!\left(\!-\frac{[(x_{p}^{i}-x_{p}^{j})+(f^{i}-f^{j})]^{2}}{2\sigma ^{2}}\!\right)\!\\
&=\exp\!\left(\!-\frac{(x_{p}^{i}-x_{p}^{j})^{2}}{2\sigma ^{2}}\!\right)\! \exp\!\left(\!-\frac{(x_{p}^{i}-x_{p}^{j})(f^{i}-f^{j})}{\sigma ^{2}}\!\right)\! \exp\!\left(\!-\frac{(f^{i}-f^{j})^{2}}{2\sigma ^{2}}\!\right)\!,\\
\end{split}
\end{equation}
where $i$, $j$ are time indices. Therefore, the new Gram matrix $K_{\mathbf{x}_p}^*$ can be represented as:
\begin{equation}\label{K_abnormal}
K_{\mathbf{x}_p}^{*}=K_{\mathbf{x}_p}\circ K_{\langle\mathbf{x}_p,~\mathbf{f}\rangle}\circ K_{\mathbf{f}}\\,
\end{equation}
where
\begin{tiny}
\begin{equation}
\begin{split}
&K_{\langle\mathbf{x}_p,~\mathbf{f}\rangle}=\\
&\begin{bmatrix}
1 &\exp\!\!\left(\!\!-\frac{(x_p^{k-w+1}-x_p^{k-w+2})(f^{k-w+1}-f^{k-w+2})}{\sigma ^{2}}\!\!\right)\!\! &\cdots & \exp\!\!\left(\!\!-\frac{(x_p^{k-w+1}-x_p^{k})(f^{k-w+1}-f^{k})}{\sigma ^{2}}\!\!\right)\!\! \\
\exp\!\!\left(\!\!-\frac{(x_p^{k-w+2}-x_p^{k-w+1})(f^{k-w+2}-f^{k-w+1})}{\sigma ^{2}}\!\!\right)\!\! &1 &\cdots & \exp\!\!\left(\!\!-\frac{(x_p^{k-w+2}-x_p^{k})(f^{k-w+2}-f^{k})}{\sigma ^{2}}\!\!\right)\!\!  \\
\vdots &\vdots &\ddots &\vdots \\
\exp\!\!\left(\!\!-\frac{(x_p^{k}-x_p^{k-w+1})(f^{k}-f^{k-w+1})}{\sigma ^{2}}\!\!\right)\!\! & \exp\!\!\left(\!\!-\frac{(x_p^{k}-x_p^{k-w+2})(f^{k}-f^{k-w+2})}{\sigma ^{2}}\!\!\right)\!\! & \cdots & 1\\
\end{bmatrix}\\
\end{split},
\end{equation}
\end{tiny}
and
\begin{equation}\label{K_fault}
\begin{split}
K_{\mathbf{f}}&=\begin{bmatrix}
   1 &\exp\!\left(\!-\frac{(f^{k-w+1}-f^{k-w+2})^{2}}{2\sigma ^{2}}\!\right)\! &\cdots & \exp\!\left(\!-\frac{(f^{k-w+1}-f^{k})^{2}}{2\sigma ^{2}}\!\right)\!  \\
   \exp\!\left(\!-\frac{(f^{k-w+2}-f^{k-w+1})^{2}}{2\sigma ^{2}}\!\right)\! & 1 &\cdots & \exp\!\left(\!-\frac{(f^{k-w+2}-f^{k})^{2}}{2\sigma ^{2}}\!\right)\!  \\
   \vdots &\vdots &\ddots & \vdots \\
   \exp\!\left(\!-\frac{(f^{k}-f^{k-w+1})^{2}}{2\sigma ^{2}}\!\right)\! & \exp\!\left(\!-\frac{(f^{k}-f^{k-w+2})^{2}}{2\sigma ^{2}}\!\right)\! &\cdots & 1\\
\end{bmatrix}\\
\end{split}.
\end{equation}

\begin{figure*}[!ht]
\setlength{\abovecaptionskip}{0.cm}
\setlength{\belowcaptionskip}{-0.0cm}
\centering
\subfigure[$K_{\mathbf{x}_1}$] {\includegraphics[width=.34\textwidth,height=3.5cm]{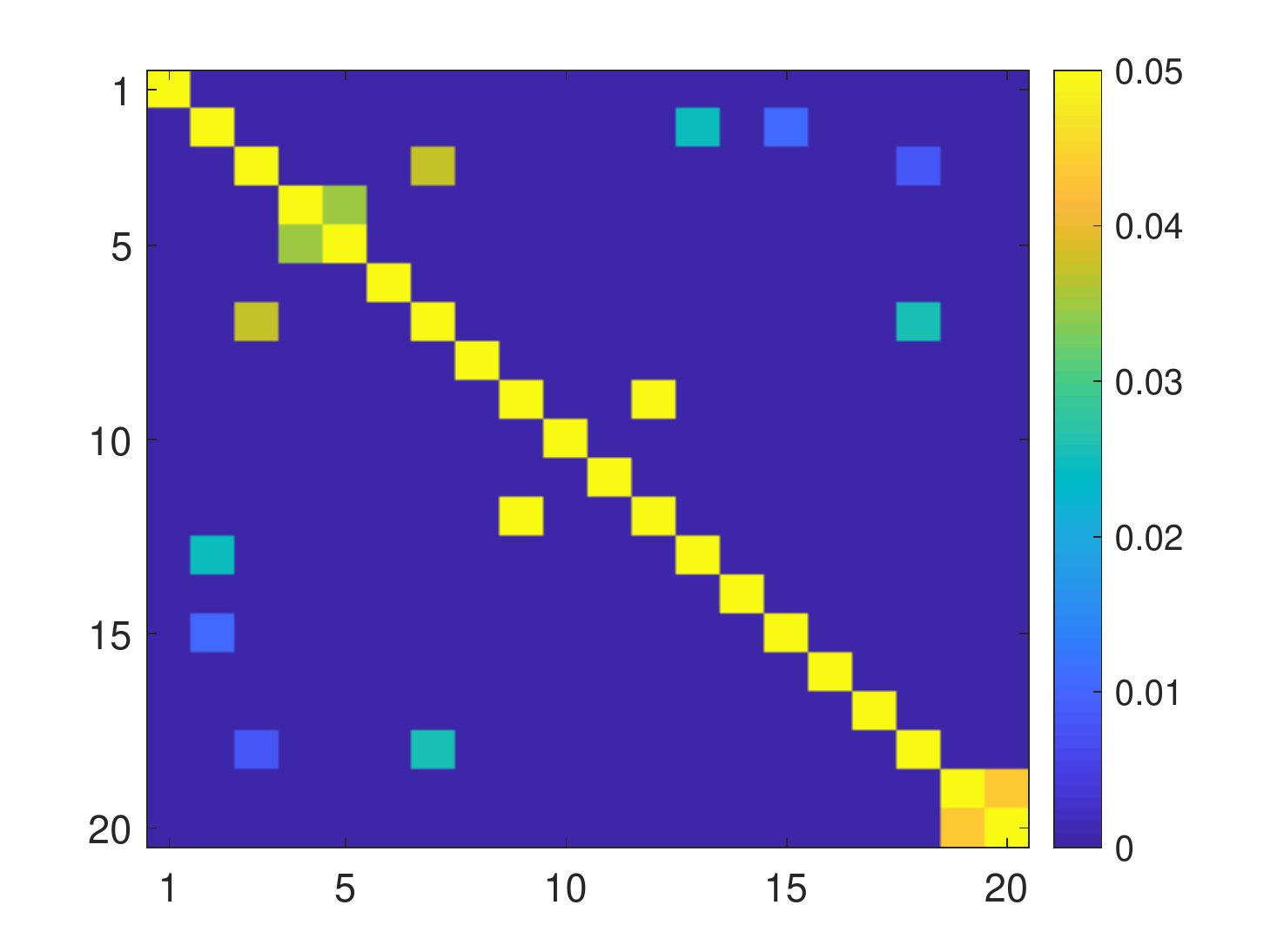}}\hspace{-2mm}
\subfigure[$K_{\mathbf{x}_1}^{*}$] {\includegraphics[width=.34\textwidth,height=3.5cm]{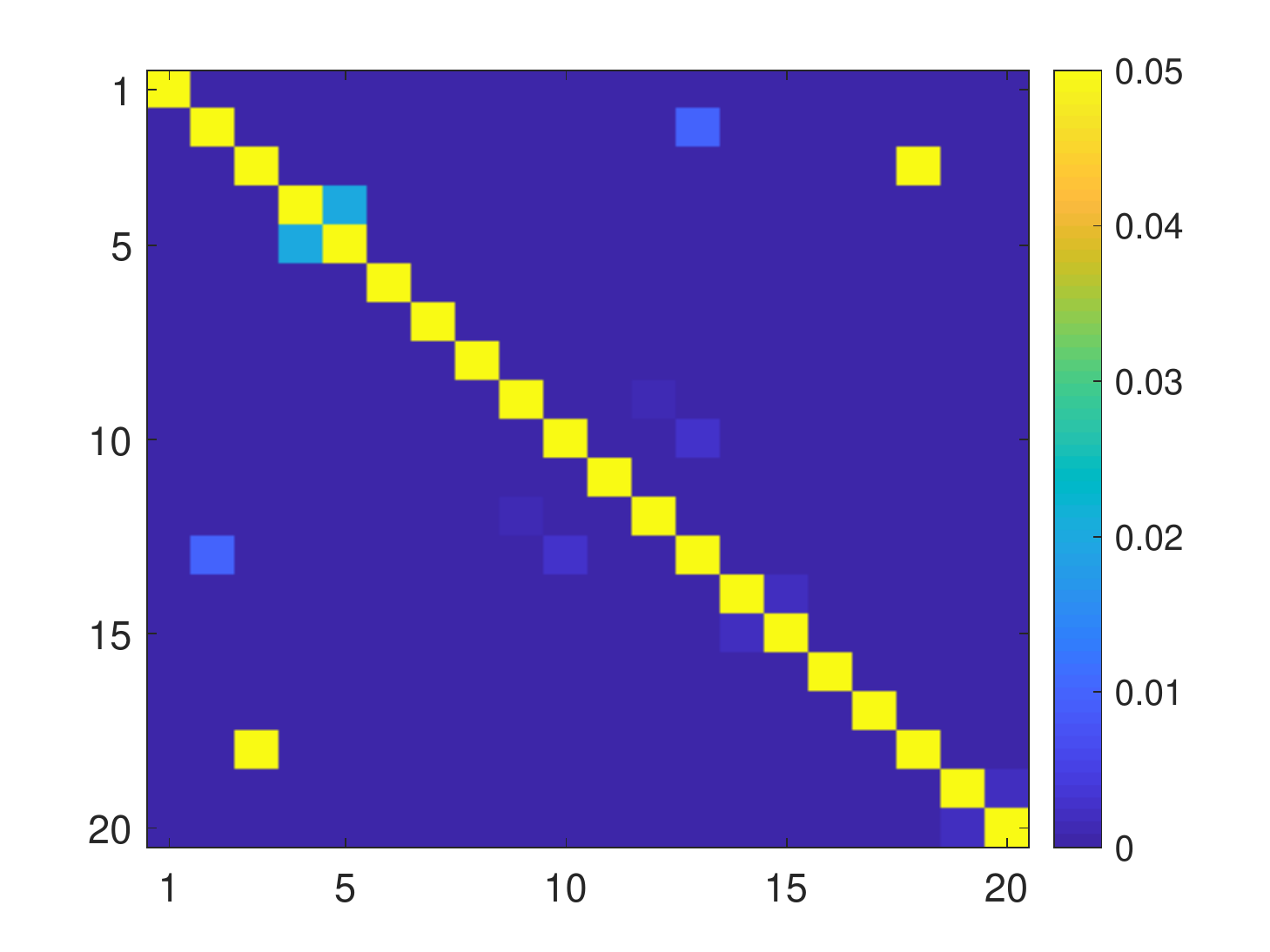}}\hspace{-2mm}
\subfigure[$Eigenvalues$] {\includegraphics[width=.32\textwidth,height=3.5cm]{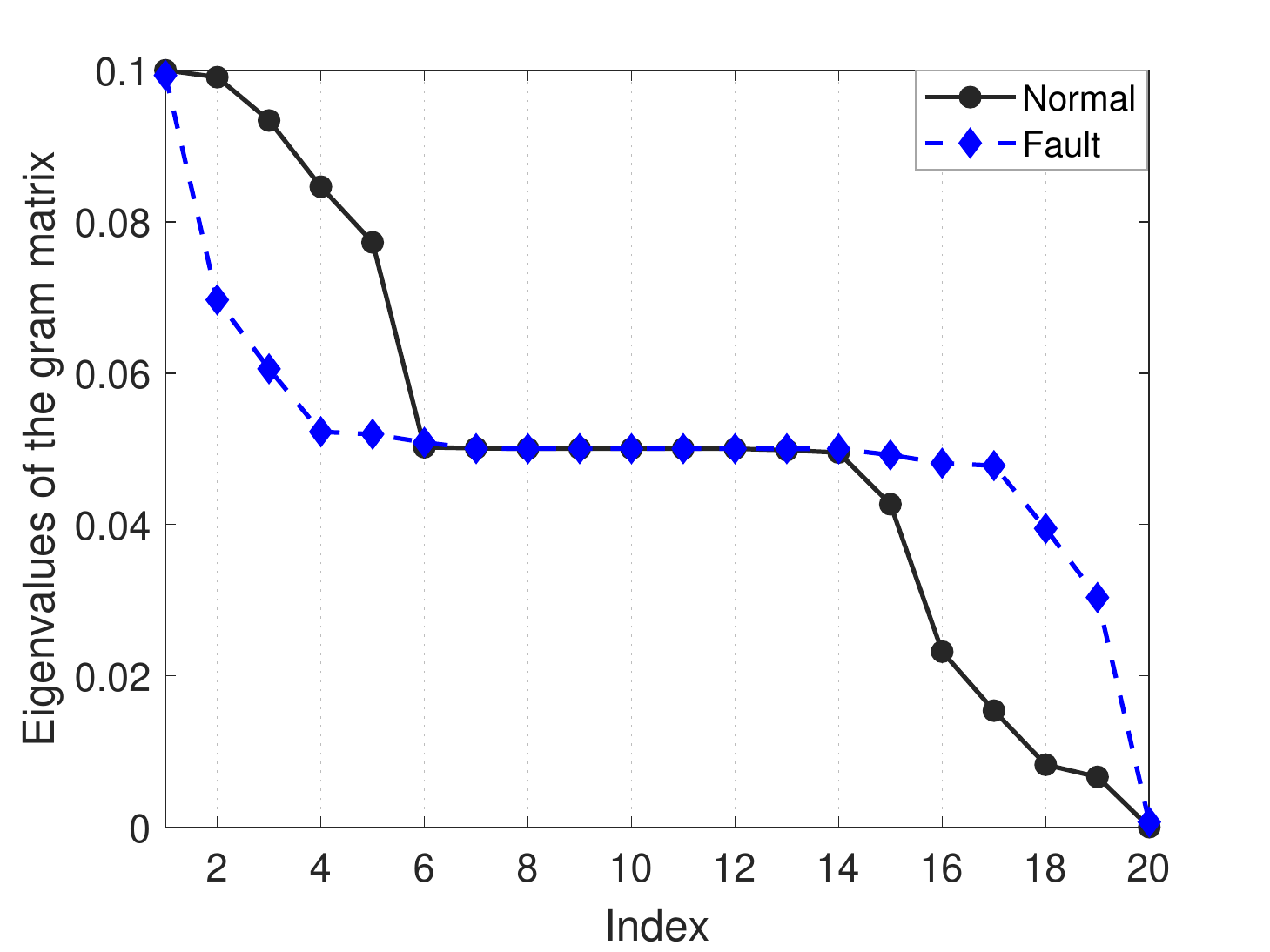}}
\caption{The (normalized) Gram matrix and its associated eigenspectrum in normal state or under incipient fault. (a) $K_{\mathbf{x}_1}$ in normal state; (b) $K_{\mathbf{x}_1}^{*}$ under incipient fault; (c) the eigenspectrum of $K_{\mathbf{x}_1}$ and $K_{\mathbf{x}_1}^{*}$. The incipient fault causes an obvious change in eigenspectrum, and thus the entropy of data.}
\label{fig:V9}
\end{figure*}

In case of incipient faults, $f^i-f^j\approx0$, Eq.~(\ref{K_fault}) reduces to an all-ones matrix. As a result, Eq.~(\ref{K_abnormal}) can be approximated with $K_{\mathbf{x}_p}^{*}\approx K_{\mathbf{x}_p}\circ K_{\langle\mathbf{x}_p,~\mathbf{f}\rangle}$.
Take the simulation data described in section 5.1 as an example, $\mathbf{f}$ is induced on $\mathbf{x}_{1}$, the Gram matrix of $\mathbf{x}_{1}$ and $\mathbf{x}_{1}^{*}$ , i.e., $K_{\mathbf{x}_1}$ and $K_{\mathbf{x}_1}^{*}$, are shown in Fig.~\ref{fig:V9}. As can be seen, the incipient fault $\mathbf{f}$ causes minor changes on the (normalized) Gram matrix as well as its eigenspectrum, and thus the entropy of the variable.

We now discuss the change of MI between the $p$-th variable $\mathbf{x}_{p}$ and the $q$-th variable $\mathbf{x}_{q}$.
Again, suppose the fault of sensor bias occurs at the $p$-th variable $\mathbf{x}_{p}^{*}$, the difference between $I(\mathbf{x}_{p};\mathbf{x}_{q})$ and $I(\mathbf{x}_p^{*};\mathbf{x}_{q})$ is:

\begin{equation}
\begin{split}
\triangle I(\mathbf{x}_p^{*};\mathbf{x}_{q})&=I(\mathbf{x}_p^{*};\mathbf{x}_{q})-I(\mathbf{x}_{p};\mathbf{x}_{q})\\
&=[H_\alpha(A_p^{*})+H_\alpha(A_{q})-H_\alpha(A_p^{*},A_{q})]-[H_\alpha(A_{p})+H_\alpha(A_{q})- H_\alpha(A_{p},A_{q})]\\
&=H_\alpha(A_p^{*})-H_\alpha(A_p^{*},A_{q})-H_\alpha(A_{p})+ H_\alpha(A_{p},A_{q})\\
&=\frac{1}{1-\alpha}\log_2 \left( \frac{\sum\limits_{i=1}^{w}\lambda_{i}(A_p^{*})^{\alpha} \sum\limits_{i=1}^{w}\lambda_{i}\left(\frac{A_p \circ A_q}{\tr(A_p \circ A_q)}\right)^{\alpha}} {\sum\limits_{i=1}^{w}\lambda_{i}(A_{p})^{\alpha}\sum\limits_{i=1}^{w}\lambda_{i}\left(\frac{A_p^* \circ A_q}{\tr(A_p^* \circ A_q)}\right)^{\alpha}} \right),\\
\end{split}
\end{equation}
where $\lambda_{i}(A)$ denotes the $i$-th eigenvalue of matrix $A$, the normalized Gram matrix obtained from the corresponding variable.

\begin{figure*}[!ht]
\setlength{\abovecaptionskip}{0.cm}
\setlength{\belowcaptionskip}{-0.0cm}
\centering
\subfigure[Normal] {\includegraphics[width=.45\textwidth]{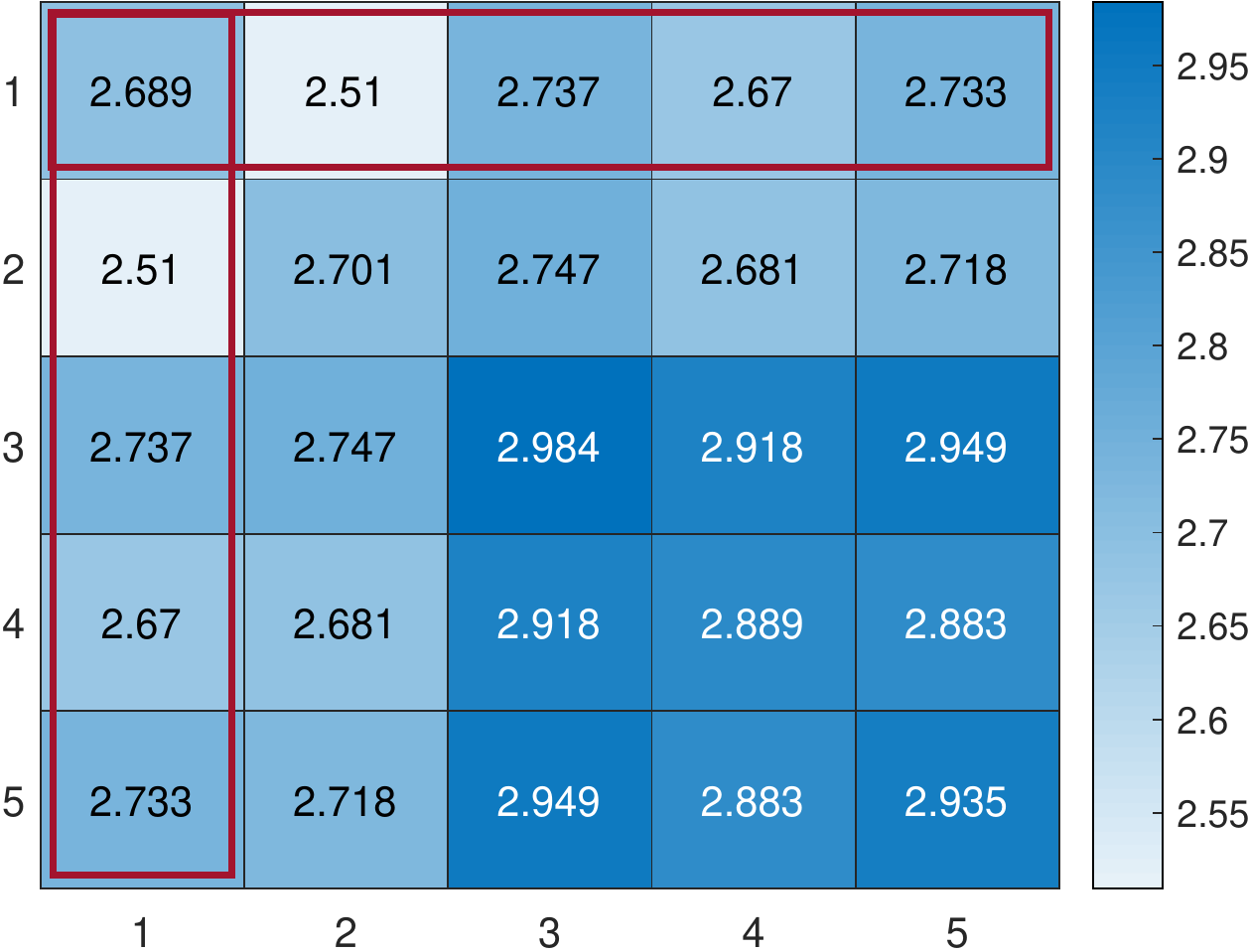}}
\subfigure[Fault] {\includegraphics[width=.45\textwidth]{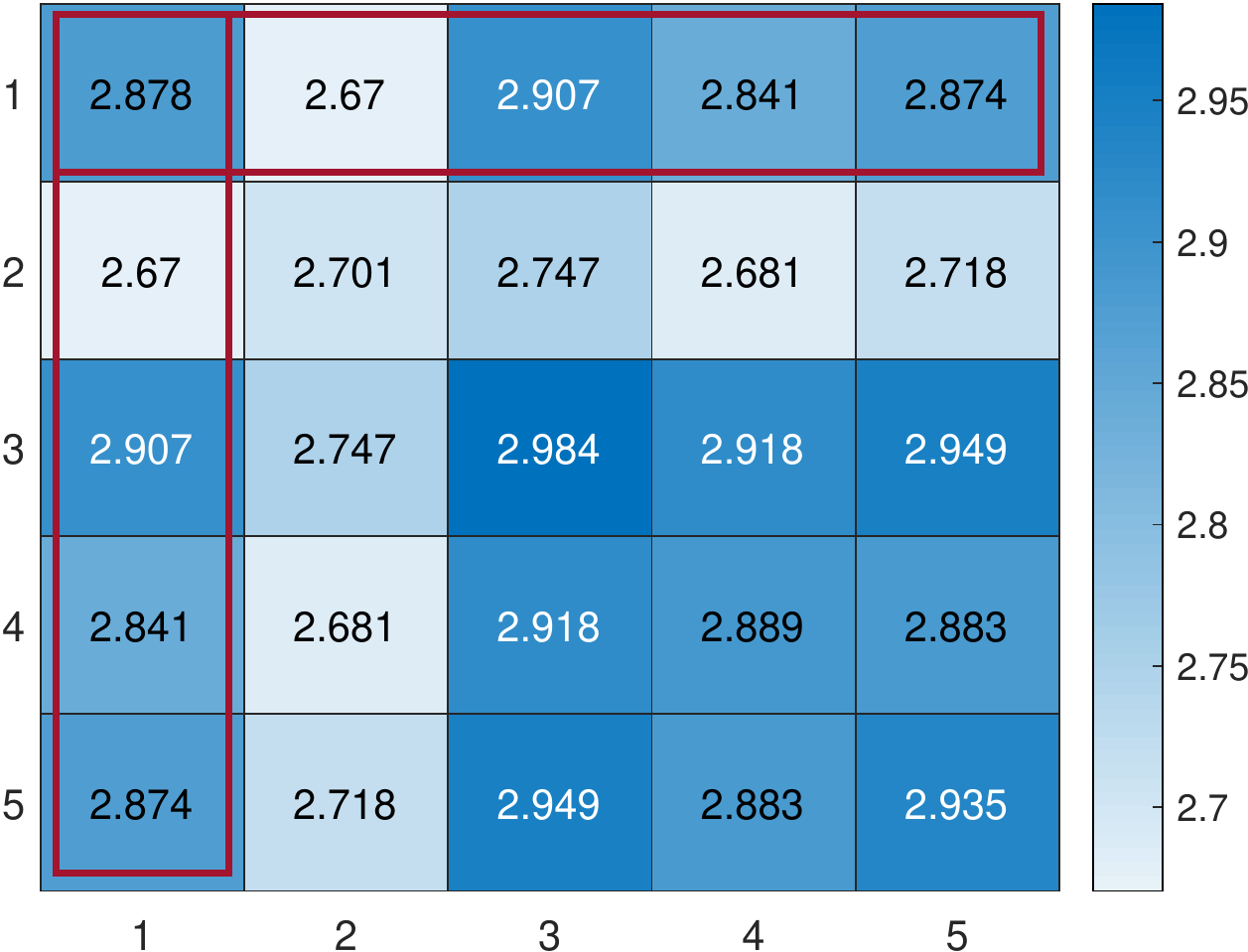}}
\caption{The MI matrix under (a) normal state; and (b) fault state (the fault is induced on $\mathbf{x}_1$). The entries with changed values are marked with red rectangles. Only entries that are related to $\mathbf{x}_1$ have different MI values.}
\label{fig:V10}
\end{figure*}

Again, we use the simulated data described in section 5.1 as an example, where the fault is induced in $\mathbf{x}_1$. By comparing the MI matrix under normal and fault states, as shown in Fig.~\ref{fig:V10}, we can observe that all entries related to $\mathbf{x}_1$ (the first dimensional measurement) have a sudden change. For example, the MI value in $M_{12}$ is $2.51$ under normal state, but it becomes $2.67$ with incipient fault. This result also indicates that our methodology has the potential to identify the exact fault sources by monitoring significant changes in MI values over MI matrix, which makes our detection result interpretable.

\section{Experiments}

In this section, experiments on both synthetic data and the real-world Tennessee Eastman process (TEP) are conducted to demonstrate the superiority of our proposed PMIM over state-of-the-art fault detection methods. We also evaluate the robustness of PMIM with respect to different hyper-parameter settings.

Two generally used metrics, namely the fault detection rate (FDR) and the false alarm rate (FAR), are employed for performance evaluation\cite{Yin1S,SXDing_2013,ZWChen_2017}.
The FDR is the probability of event where an alarm is raised when a fault really occurs,
\begin{equation}
\text{FDR}=\text{prob}(D>D_{\text{cl}}|\text{fault $\neq$ 0}), \\
\end{equation}
where $D$ and $D_{\text{cl}}$ are respectively the similarity index and its corresponding control limit.
By contrast, the FAR is the percentage of the samples under normal state but are identified as faults,
\begin{equation}
\text{FAR}=\text{prob}(D>D_{\text{cl}}|\text{fault = 0}). \\
\end{equation}
Obviously, a higher FDR and a lower FAR is expected.

\subsection{Numerical Simulation}

Motivated by \cite{Alcala_2009,Shang_2017,Shang_2018}, we consider a multivariate nonlinear process generated by the following equation:
\begin{equation} \nonumber
\begin{bmatrix}
   x_{1}  \\
   x_{2}  \\
   x_{3}  \\
   x_{4}  \\
   x_{5}  \\
\end{bmatrix}=\begin{bmatrix}
 {0.2183} & { - 0.1693} & {0.2063}  \\
   { - 0.1972} & {0.2376} & {0.1736}  \\
   {0.9037} & { - 0.1530} & {0.6373}  \\
   {0.1146} & {0.9528} & { - 0.2624}  \\
   {0.4173} & { - 0.2458} & {0.8325}  \\
\end{bmatrix} \begin{bmatrix}
   {s_{1}}^2  \\
   s_{2}s_{3}  \\
   {s_{3}}^3  \\
\end{bmatrix}+\begin{bmatrix}
   e_{1}  \\
   e_{2}  \\
   e_{3}  \\
   e_{4}  \\
   e_{5}  \\
\end{bmatrix},
\end{equation}
where $s$ satisfies $s^k_i=\sum_{j=1}^{l}\beta_{i,j}v^{k-j+1}_i$ with a weight matrix $\mathbf{\beta}$ given by,
\begin{equation} \nonumber
\mathbf{\beta}=\begin{bmatrix}
   {0.6699} & {0.0812} & {0.5308} & {0.4527} & {0.2931}  \\
   {0.4071} & {0.8758} & {0.2158} & { - 0.0902} & {0.1122}  \\
   {0.3035} & {0.5675} & {0.3064} & {0.1316} & {0.6889}  \\
\end{bmatrix},
\end{equation}
$v$ denotes three mutually independent Gaussian distributed data sources with mean of $[0.3,~2.0,~3.1]^{T}$ and standard deviation of $[1.0,~2.0,~0.8]^{T}$, and $e$ denotes Gaussian white noises with standard deviation
$[0.061,~0.063,~0.198,~0.176,~0.170]^{T}$. Same to \cite{Shang_2017,Shang_2018}, we consider four different types of faults that cover a broad spectrum of real-life scenarios,
\begin{itemize}
    \item Type I: Sensor bias $\mathbf{x}^{*}=\mathbf{x}+f$, with $f=5.6+\mathbf{e}$, $\mathbf{e}$ randomly chosen from [0,~1.0];
    \item Type II: Sensor precision degradation $\mathbf{x}^{*}=\eta \mathbf{x}$ with $\eta=0.6$;
    \item Type III: Additive process fault $\mathbf{s}^{*}=\mathbf{s}+f$ with $f=1.2$;
    \item Type IV: Dynamic changes $\mathbf{\tilde{\beta}}=\mathbf{\beta}+\bigtriangleup \mathbf{\beta}$ with $\bigtriangleup \beta_{3}=[-0.825,~0.061,~0.662,~-0.820,
~0.835]$, where $\mathbf{\beta}_{3}$ denotes the $3$-th row of $\mathbf{\beta}$.
\end{itemize}

The training set contains $10,000$ samples, the test set contains $4,000$ samples. All the faults are introduced after the $1,000$-th sample. For convenience, we assume sensor fault occurs at $\mathbf{x}_{1}$ (i.e., the first dimension of observable measurement), and process fault occurs at $\mathbf{s}_{1}$ (i.e., the first independent data sources). Empirical evaluation aims to answer the following three questions:

\begin{itemize}
\item Can MI manifest more complex dependence among different dimensions of measurement than the classical correlation coefficient?
\item Is fault detection using PMIM robust to hyper-parameter settings and how hyper-parameters affect the performance of PMIM?
\item Does PMIM outperform existing state-of-the-art window-based fault detection methods?
\end{itemize}

\subsubsection{MI versus Pearson's correlation coefficient}

Firstly, we demonstrate the advantage of MI over the Pearson's correlation coefficient $\gamma$ on manifesting the complex (especially nonlinear) dependency between two variables. Intuitively, if two random variables are linearly correlated, they should have large $\gamma^{2}$ ($\gamma^{2}> 0.6$) and large MI\footnote{In general, $\gamma^{2}>0.3$ indicates a moderate linear dependence and $\gamma^{2}>0.6$ indicates a strong linear dependence~\cite{ratner2009correlation,Jiang_2011}. However, there is little guidance for what value of MI really constitutes an indication of strong dependence~\cite{Jiang_2011}. This is just because MI is not upper bounded and different estimators usually offer different MI values. Therefore, we intuitively consider a MI value is ``large" if the corresponding $\gamma^2$ indicates a ``strong" linear dependence (i.e., larger than $0.6$).}(but we cannot compare the value of $\gamma^2$ to the value of MI).
However, if they are related in a nonlinear fashion, they should have large MI but small $\gamma^{2}$ ($\gamma^{2}\leq 0.6$)~\cite{Jiang_2011}. On the other hand, two variables will never have a large $\gamma^2$ but a small MI, as linear correlation is a very special case of the general dependence. Therefore, MI should always be a superior metric to measure the degree of interactions than Pearson's correlation coefficient. We perform a simple simulation to support our argument.
\begin{figure*}[!t]
\setlength{\abovecaptionskip}{0.cm}
\setlength{\belowcaptionskip}{-0.0cm}
\centering
\subfigure[Shannon's MI versus $\gamma^{2}$] {\includegraphics[width=.5\textwidth]{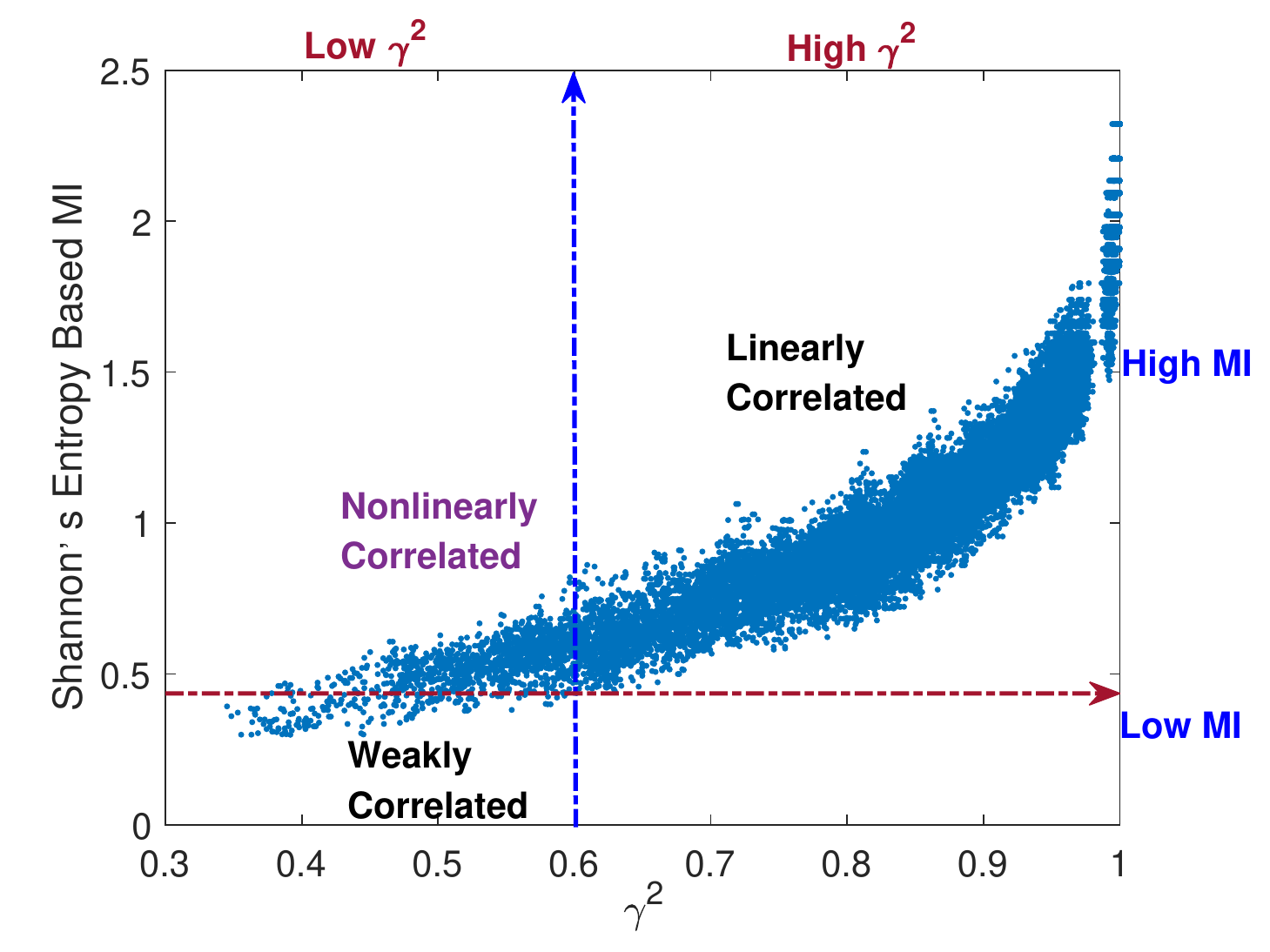}}\hspace{-4mm}
\subfigure[Matrix-based R{\'e}nyi's $\alpha$-order MI versus $\gamma^{2}$] {\includegraphics[width=.5\textwidth]{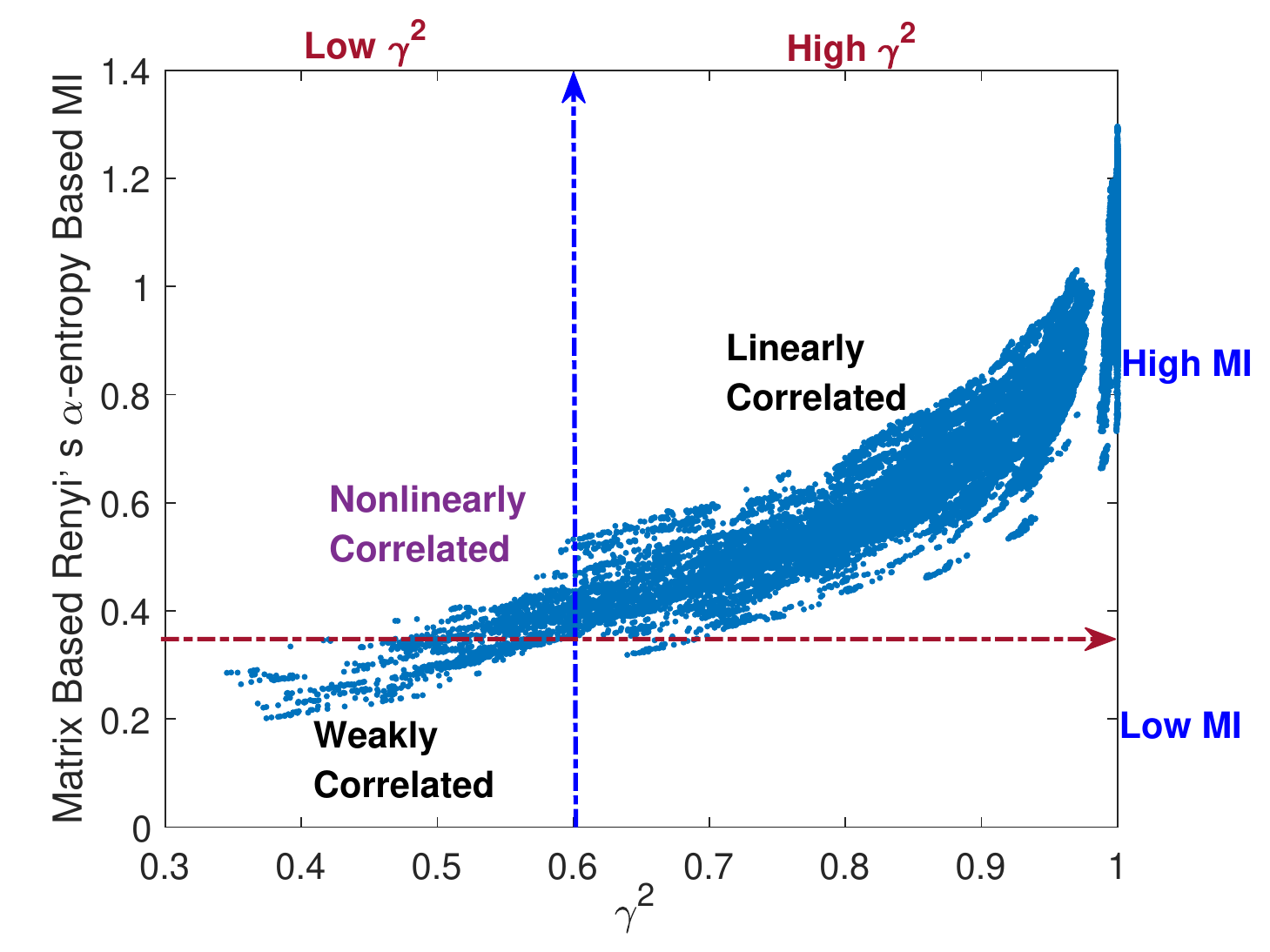}}
\caption{The comparison between Pearson's correlation coefficient $\gamma^{2}$ and mutual information estimated with (a) Shannon's discrete entropy functional by discretizing continuous variables into $5$ bins of equal width; and (b) matrix-based R{\'e}nyi's $\alpha$-order MI. The values of $\gamma^{2}$ and MI are shown in $x$-axis and $y$-axis, respectively.}\label{fig:PearsonMI}
\end{figure*}

Specifically, we select the first $4,000$ samples in the training set and compute both MI and $\gamma^{2}$ in each window data of size $100$. We finally obtain $3,601$ pairs of MI and $\gamma^{2}$.
We evaluate MI with both the basic Shannon's discrete entropy functional and our suggested matrix-based R{\'e}nyi's $\alpha$-order entropy functional. For Shannon entropy functional, we discretize continuous variables into $5$ bins of equal width to estimate the underlying distributions.
The values of MI ($y$-axis) and $\gamma^{2}$ ($x$-axis) are specified in the scatter plot in Fig.~\ref{fig:PearsonMI}. As can be seen, there are strong nonlinear dependencies in our simulated data. Take Fig.~\ref{fig:PearsonMI}(b) as an example, we can observe that when $\gamma^{2}=0.6$, the smallest MI is $0.37$.
As such, we consider $MI\geq0.37$ to indicate a strong correlation. We noticed that there are quite a few points in the region $0.37\leq MI \leq 1.2$ and $\gamma^{2}\leq 0.6$, suggesting that nonlinear dependence dominates for a large number of variables.

%This indicates that there is strong (linear) dependence in the area of $MI\in [0.37,~1.2]$, and $\gamma^{2}\in [0.6,~1.0]$. However, the existence of multiple points in the area $MI\in [0.37,~1.2]$ and $\gamma^{2}\in [0.3,~ 0.6]$ suggests that some variables contain strong nonlinear dependencies that can only be quantified by MI. Moreover, one should note that, there are almost no points in the area that $MI\in [0,~ 0.37]$ and $\gamma^{2}\in [0.6,~1.0]$. In other words, all dependencies detected by correlation coefficient can also be captured by MI.

\begin{figure*}[!t]
\setlength{\abovecaptionskip}{0.cm}
\setlength{\belowcaptionskip}{-0.0cm}
\centering
\subfigure[FDRs] {\includegraphics[width=.5\textwidth]{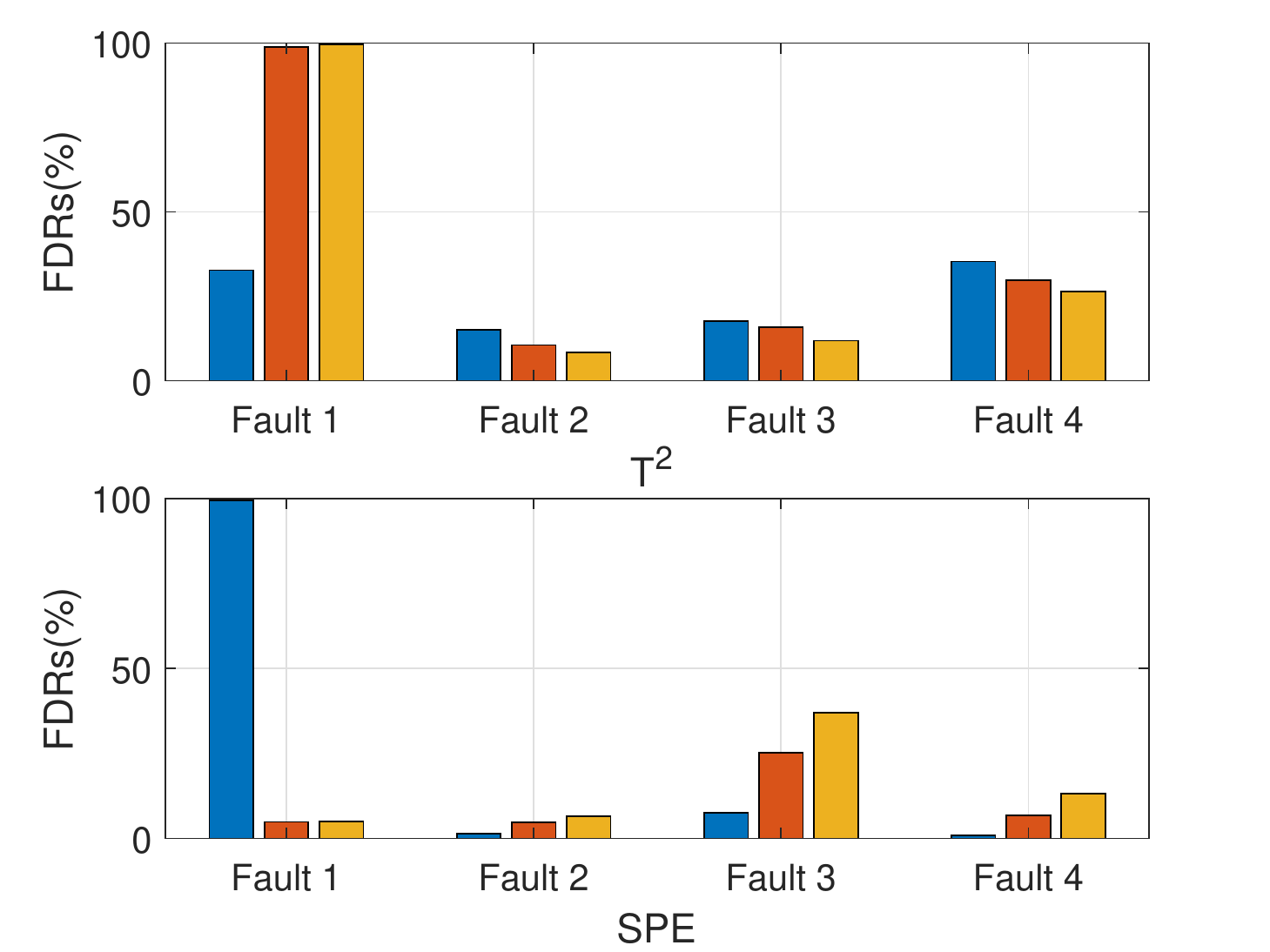}}\hspace{-4mm}
\subfigure[FARs] {\includegraphics[width=.5\textwidth]{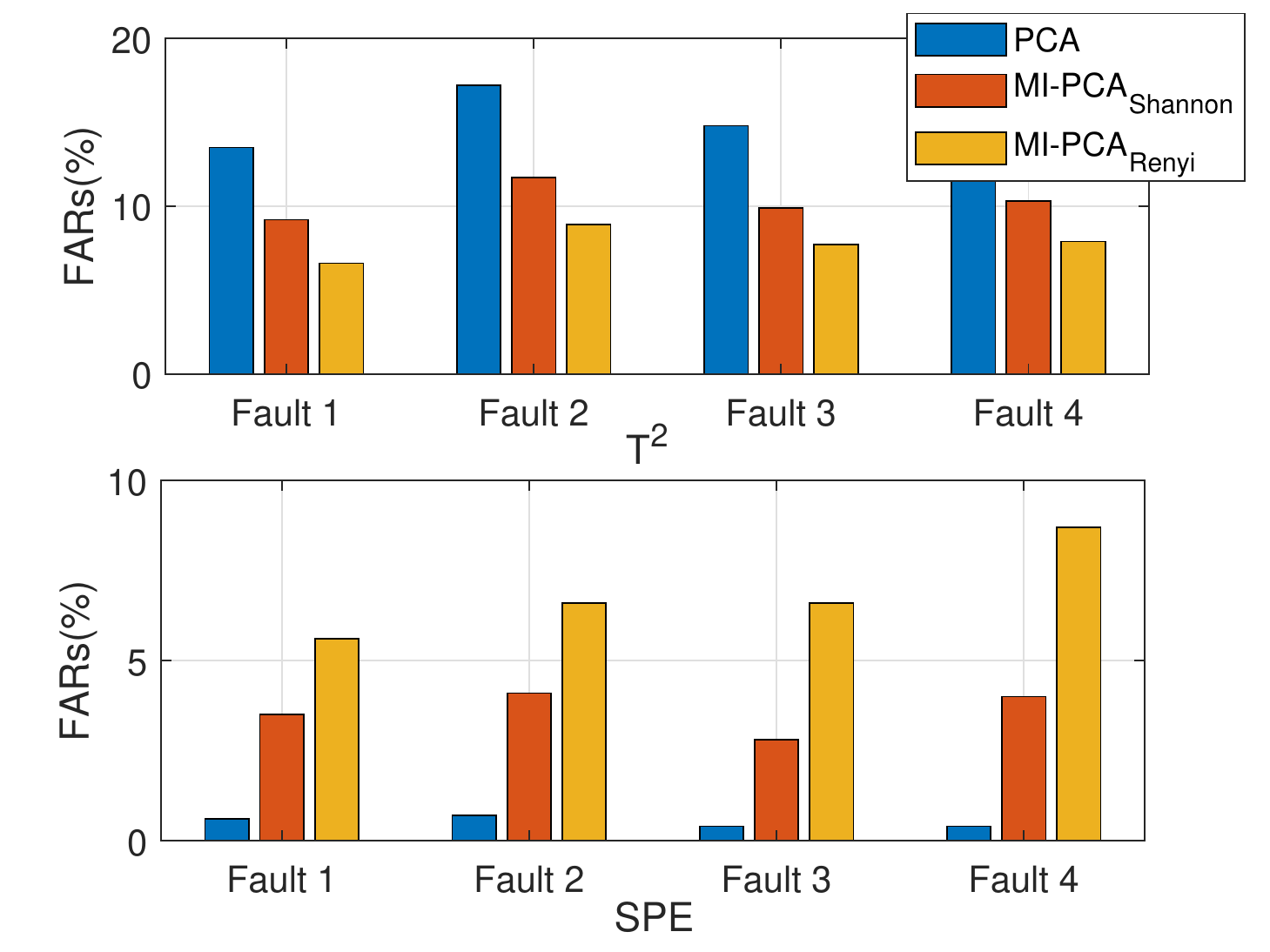}}
\caption{Performance comparison between PCA and MI-PCA in terms of FDR (the larger the better) and FAR (the smaller the better). We replace the covariance matrix in the basic PCA-based fault detection with MI matrix estimated with both Shannon entropy (denote it MI-PCA$_{\text{Shannon}}$) and matrix based R{\'e}nyi's $\alpha$-order entropy (denote it MI-PCA$_{\text{R{\'e}nyi}}$).
We use both Hotelling $T^2$ and squared prediction error (SPE) to monitor the state of samples.}
\label{fig:PCAMI}
\end{figure*}

Further, to quantitatively demonstrate the superiority of MI matrix over the well-known covariance matrix on nonlinear fault detection, we use MI matrix as a substitute to the covariance matrix in the basic PCA-based fault detection approach. We denote this simple modification as MI-PCA, which includes both MI-PCA$_{\text{Shannon}}$ and MI-PCA$_{\text{R{\'e}nyi}}$. Both Hotelling $T^2$ and squared prediction error (SPE) are considered in PCA and MI-PCA. Performances in terms of FDR and FAR are shown in Fig.~\ref{fig:PCAMI}. In case of $T^2$, MI-PCA always has higher or almost the same FDR values, but significantly smaller FAR values. In case of SPE, although traditional PCA has smaller FAR, its results are meaningless. In fact, if we look deeper, the FDR of PCA is almost zero, which suggests that traditional PCA completely fails.

%\textcolor{red}{When detecting the faults induced here, samples belonging to normal and abnormal conditions may overlap to a large extent owing to their small magnitudes\footnote{\textcolor{blue}{Take fault 1 as an example, the standard deviation of $\mathbf{x}_{1}$ changes from 19.05 to 19.15 after fault induced.}}. However, PCA always neglect the temporal correlation between consecutive samples .}

\subsubsection{Hyperparameter analysis}

\begin{figure*}[!t]
\setlength{\abovecaptionskip}{0.cm}
\setlength{\belowcaptionskip}{-0.0cm}
\centering
\subfigure[FDRs~with~different~$\alpha$] {\includegraphics[width=.5\textwidth]{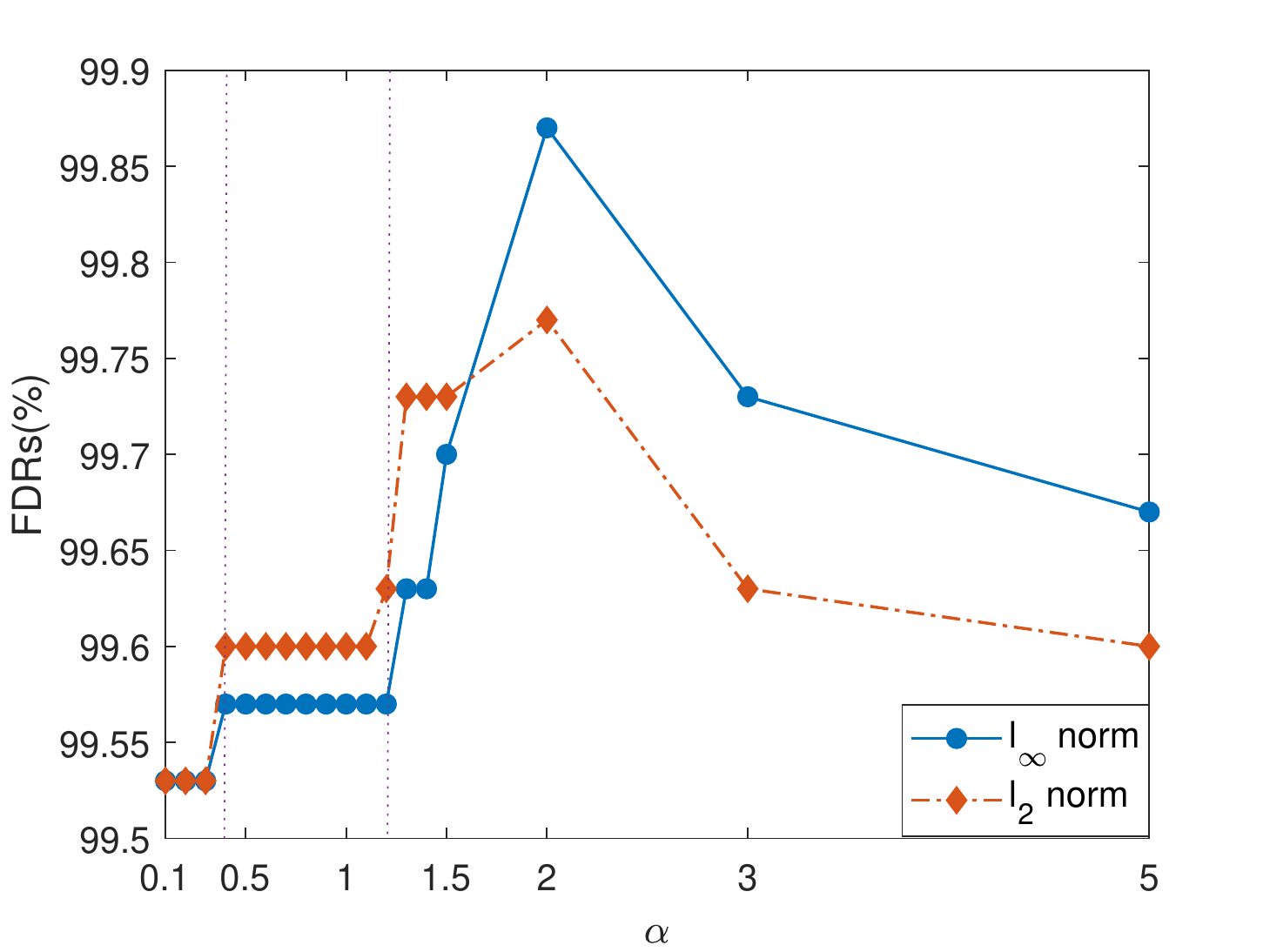}}\hspace{-4mm}
\subfigure[FARs~with~different~$\alpha$] {\includegraphics[width=.5\textwidth]{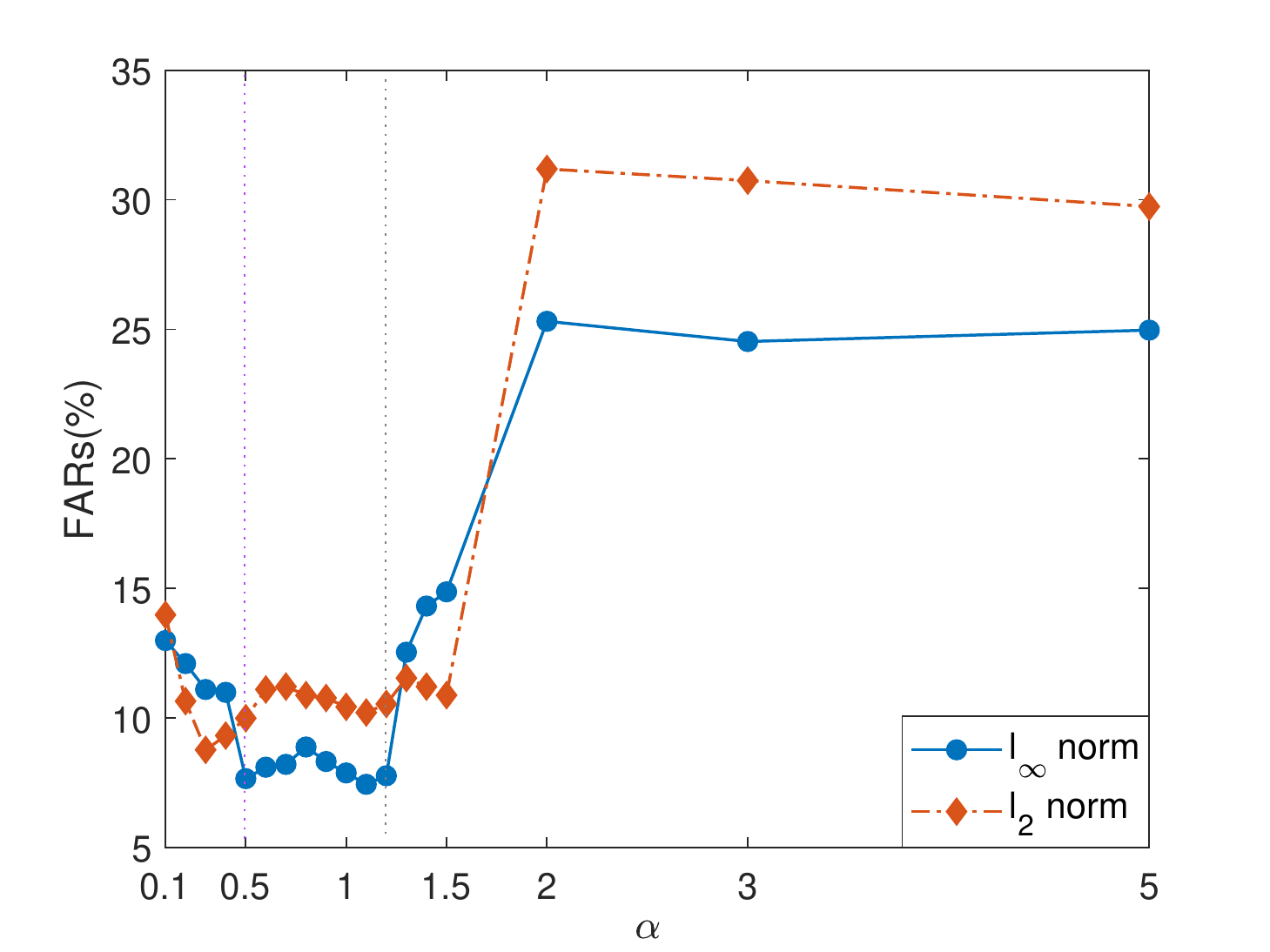}}
\caption{Detection performances of different $\alpha$ on (a) FDRs; and (b) FARs. Both $\ell_{\infty}$ and $\ell_{2}$ norm are considered in the calculation of similarity index $D$. As a common practice, window size $100$ is used here.}
\label{fig:FDR}
\end{figure*}

\begin{figure*} [!ht]
\setlength{\abovecaptionskip}{0.cm}
\setlength{\belowcaptionskip}{-0.0cm}
\centering
\subfigure[FDRs~with~different~$\sigma$] {\includegraphics[width=.5\textwidth]{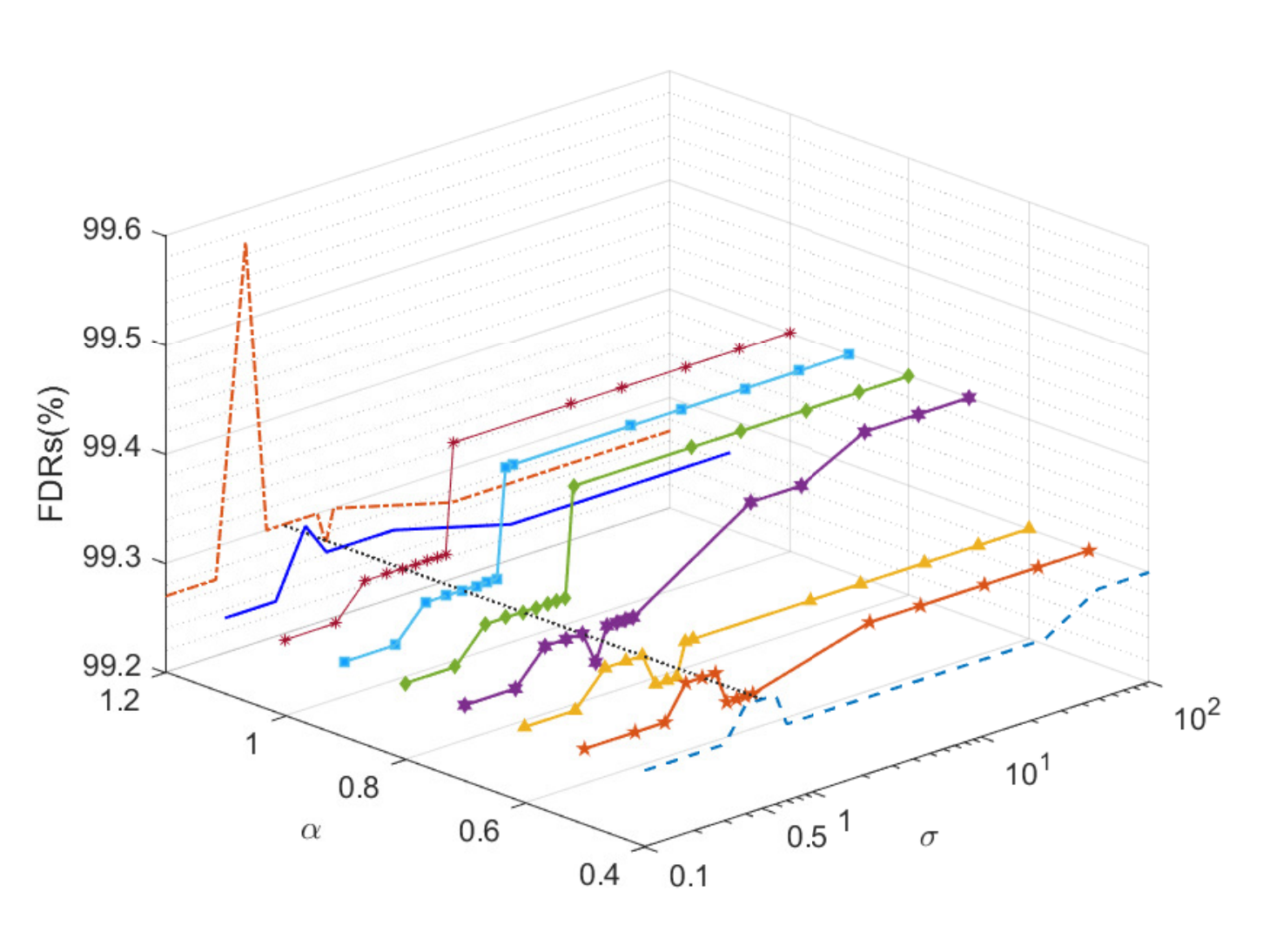}}\hspace{-4mm}
\subfigure[FARs~with~different~$\sigma$] {\includegraphics[width=.5\textwidth]{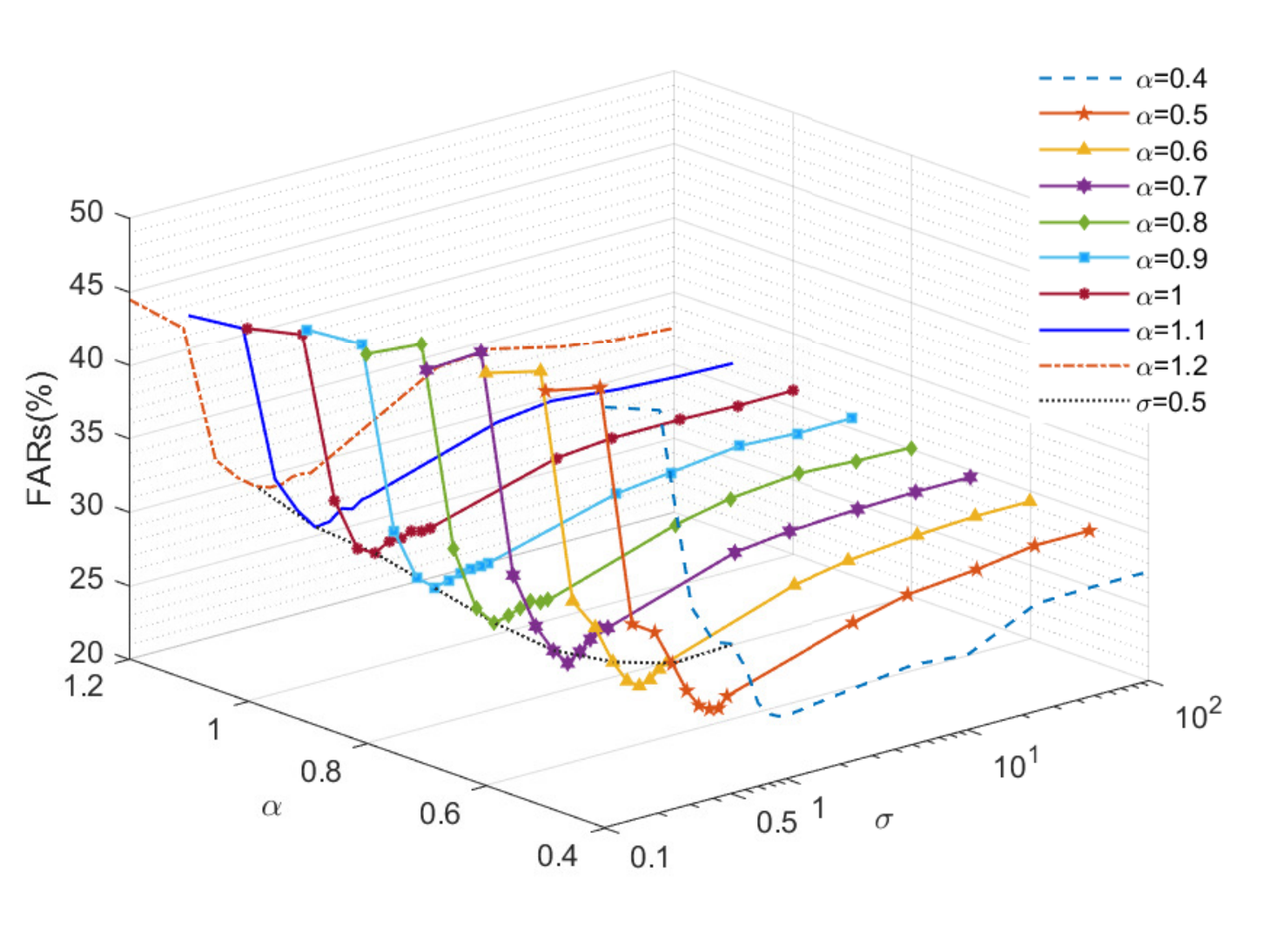}}
\caption{Detection performances of different $\sigma$ with a fixed $\alpha$ on (a) FDRs; and (b) FARs. $\sigma\in\{0.1,~0.2,~0.3,~0.4,$ $~0.5,~0.6,~0.7,~0.8,$ $~0.9,~1,~5,~10,~24,~50,~100\}$ (displayed in log-scale). $\ell_{2}$ norm is considered in the calculation of similarity index $D$.}
\label{fig:FDRFAR}
\end{figure*}

\begin{figure*} [!ht]
\setlength{\abovecaptionskip}{0.cm}
\setlength{\belowcaptionskip}{-0.0cm}
\centering
\subfigure[FDRs~with~different~$w$] {\includegraphics[width=.51\textwidth]{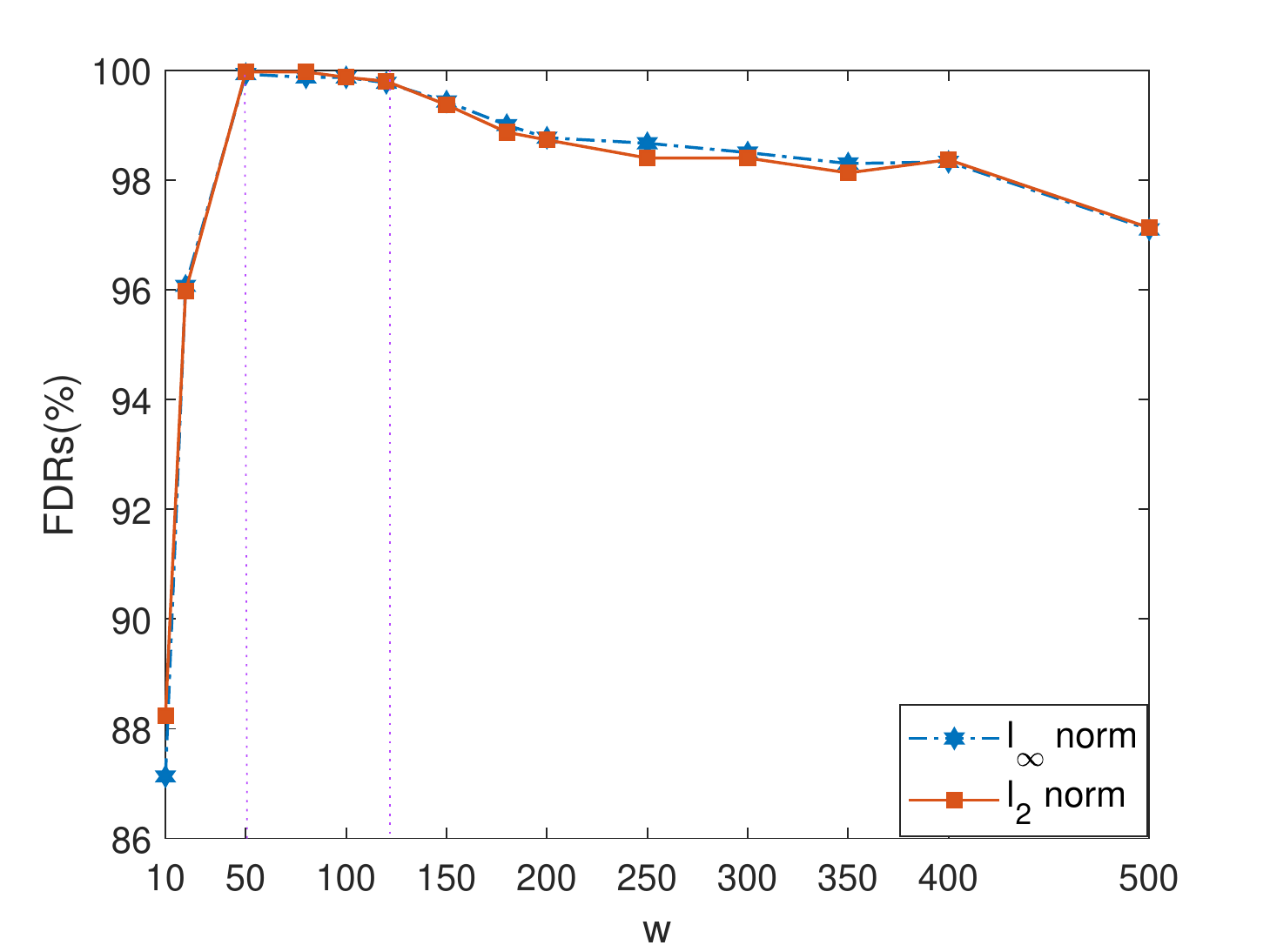}}\hspace{-4mm}
\subfigure[FARs~with~different~$w$] {\includegraphics[width=.51\textwidth]{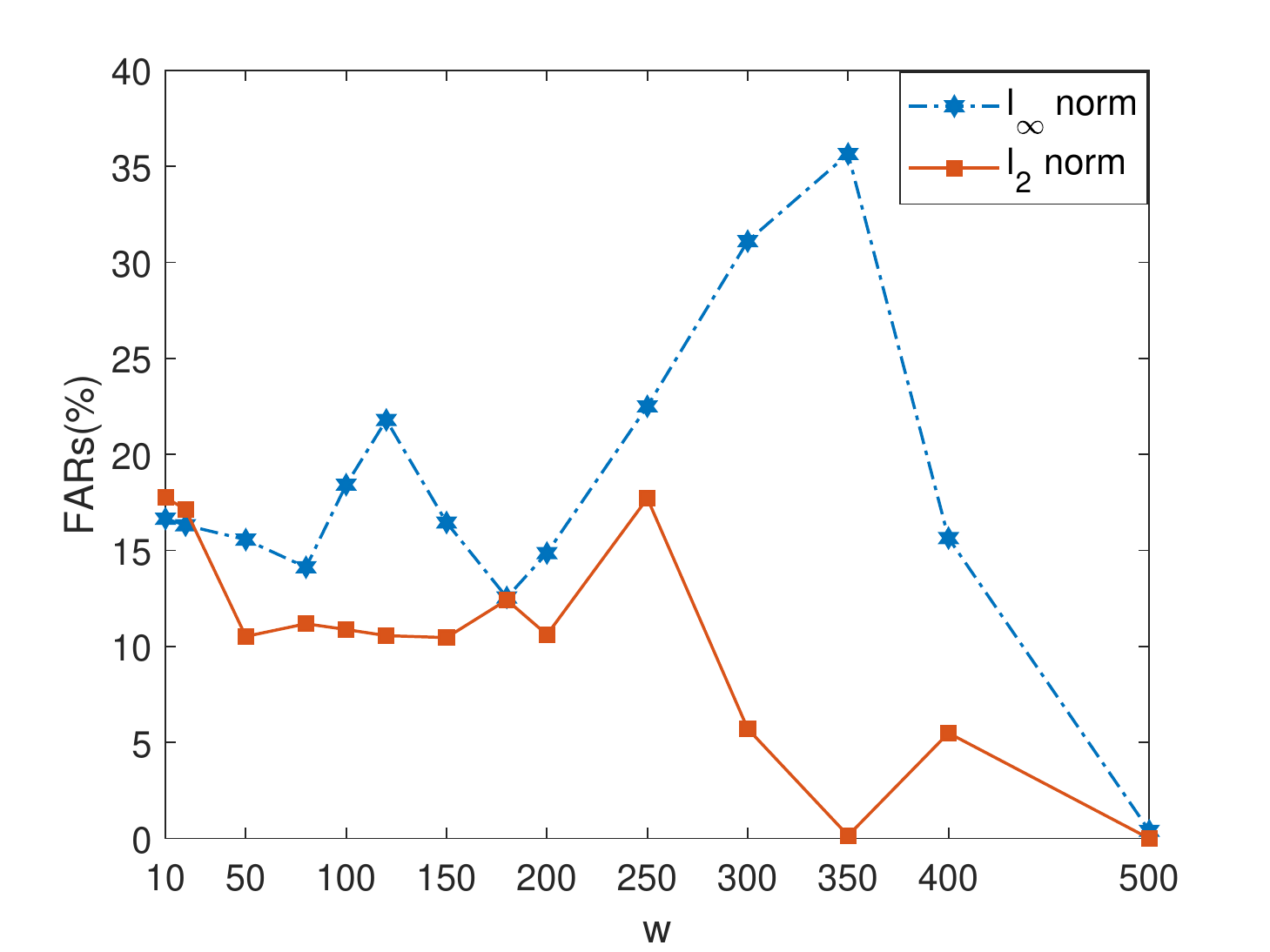}}
\caption{Detection performances of different $w$ on (a) FDRs; and (b) FARs. Both $\ell_{\infty}$ and $\ell_{2}$ norm are considered for scalarization in the calculation of similarity index $D$.}
\label{fig:FAR}
\end{figure*}

We then present a comprehensive analysis on the effects of three hyper-parameters, namely the entropy order $\alpha$, the kernel size $\sigma$ and the length $w$ of sliding window in PMIM. We focus our discussion on the process data with time-correlated dynamic changes, i.e., fault Type V. The FDR and FAR values of our methodology with respect to different hyper-parameter settings are shown in Fig.~\ref{fig:FDR}, Fig.~\ref{fig:FDRFAR} and Fig.~\ref{fig:FAR}.

The choice of $\alpha$ is associated with the task goal. If the application requires emphasis on tails of the distribution (rare events) or multiple modalities, $\alpha$ should be less than 2, but if the goal is to characterize modal behavior, $\alpha$ should be greater than 2. $\alpha= 2$ provides neutral weighting \cite{yu2019multivariate,yu2019simple}. The detection performances of different values of $\alpha$ are presented in Fig.~\ref{fig:FDR}. For a comprehensive comparison, we consider $\alpha\in\{0.1,~0.2,~0.3,~0.4,~0.5,~0.6,~0.7,~0.8,~0.9,~1,$ $~1.1,~1.2,~1.3,~1.4,~1.5,~2,~3,~5\}$. Both $\ell_{\infty}$ and $\ell_{2}$ are assessed in the calculation of similarity index $D$ in Eq.~(\ref{similarity}). As a common practice, we use window size $100$. As can be seen, the FDR values are always larger than $99.5\%$, which suggests that FDR is less sensitive to the changes of $\alpha$. On the other hand, the FAR keeps a stable value in the range $\alpha \in[0.5,~1.2]$, but suddenly increases to $25\%$ or above when $\alpha \geq 2$. Therefore, we recommend $\alpha$ in the range $[0.5,~1.2]$ for PMIM.

The parameter $\sigma$ controls the locality of the estimator, its selection can follow Silverman's rule of thumb for density estimation~\cite{Silverman_1986} or other heuristics from a graph cut perspective (e.g., the $10$ to $30$ percent of the total range of the Euclidean distances between all pairwise data points~\cite{shi2000normalized}). For example, the range from a graph cut perspective corresponds to $0.21 < \sigma< 1.33$ on the normalized data here. The detection performances of different $\sigma$ and $\alpha$ are presented in Fig.~\ref{fig:FDRFAR}. We choose $\sigma\in\{0.1,~0.2,~0.3,~0.4,~0.5,~0.6,$ $~0.7,~0.8,~0.9,~1,~5,~10,~24,~50,~100\}$ (displayed in log-scale) and $\alpha\in\{0.4,~0.5,$ $~0.6,~0.7,~0.8,~0.9,~1,~1.1,~1.2,~1.5\}$.
According to Fig.~\ref{fig:FDRFAR}, FDR is always larger than $99.20\%$, whereas FAR is relatively more sensitive to $\sigma$.
Specifically, FAR reaches to its minimum value when $\sigma$ is around $0.5$. After that, FAR is consistently increasing when $\sigma \in[1,~100]$. To achieve higher FDR and lower FAR values, we thus recommend $\sigma$ in the range $[0.4,~1]$ for PMIM.

%Increasing the size of sliding window reduces the variations in corresponding entries in consecutive MI matrices, thus is prone to produce a more stable distribution of their eigenspectrum. \textcolor{blue}{However, a smaller sliding window $w$ would break the stationarity assumption in our methodology. If the window length is too small, the changes of the MI values triggered by the fault will be overwhelmed and the trajectory of the fault will be difficult to statistic in the transition phase, which leads to poor fault detection capability and an unacceptable detection delay.}

The local stationarity or smoothness assumption (of the underlying process) might be violated if the window size is too large. In this case, the eigenspectrum becomes stable and is less sensitive to the abrupt distributional changes of the underlying process, which may lead to decreased detection power or lower FDR values. On the other hand, in case of a very small window size, the MI estimation becomes unreliable (due to limited samples) and the local time-lagged matrix may be dominated by environmental noises, which in turn would result in a high FAR value. Moreover, according to Fig.~\ref{fig:FAR}, FDR remains stable when $w\in [50,~120]$, and decreases as the window length increasing when $w \geq 120$. By contrast, FAR is more sensitive to $w$ than FDR, but its changing patterns are not consistent for $\ell_2$ norm and $\ell_\infty$ norm. We choose $w=100$ in the following experiments, because it can strike a good trade-off between FDR and FAR for both $\ell_2$ norm and $\ell_\infty$ norm here.

\subsubsection{Comparison with state-of-the-art methods}

We compare our proposed PMIM with four state-of-the-art window based data-driven fault detection approaches, namely DPCA \cite{Ku_1995}, SPA \cite{Wang_2010}, RTCSA \cite{Shang_2017} and RDTCSA \cite{Shang_2018}. The hyperparameters of PMIM are set to $\alpha=1.01$, $\sigma=0.5$ and $w=100$.
%here \footnote{Although $w=80$ has been verified friendly to MI-TCSA method, our method can still achieve better results under commonly used sizes, such as 100, 120, 150. See Appendix}.
For DPCA, $90\%$ cumulative percent variance is used to determine the number of principal components. For RTCSA, RDTCSA and PMIM, their detection performances are illustrated in Table~\ref{1:TableFDR} and Table~\ref{1:TableFAR}.
{
\linespread{1}
\begin{table}[!hbpt]
\small
\centering
\begin{threeparttable}
\caption{The FDRs $(\%)$ of different methods for the numerical simulations} \label{1:TableFDR}
\renewcommand{\arraystretch}{1.1}
\renewcommand{\tabcolsep}{2mm}
\begin{tabular}{c||c|c||c|c||c||c||c}\hline
\toprule
\textbf{No.}&  \multicolumn{2}{|c||}{\textbf{DPCA}}
&\multicolumn{2}{|c||}{\textbf{SPA}} &\textbf{RTCSA} &\textbf{RDTCSA} &\textbf{PMIM} \\ \cline{2-3}\cline{4-5}
&\footnotesize{$T^{2}$} &\footnotesize{SPE} &\footnotesize{$D_{r}$} &\footnotesize{$D_{p}$}  &  &  &  \\\hline
\midrule
  1 &51.17  &\textbf{99.70}   &0.80 &2.80 &88.43 &91.01  &91.57 \\  \hline
  2 &21.23  &21.0 &2.40 &6.67 &82.50 &\textbf{100}  &99.63  \\  \hline
  3 &33.10  &\textbf{99.83} &0.77 &7.37 &96.60 &96.83  &97.50   \\  \hline
  4 &81.23  &85.57 &29.13 &99.13 &99.70 &99.70  &\textbf{99.87}   \\  \hline
  \textbf{Aver.} &46.68  &76.53 &8.28 &29.0 &91.81  &96.89 &\textbf{97.14}  \\  \hline
\bottomrule
\end{tabular}
\begin{tablenotes}
\footnotesize
\item $T^2$ denotes Hotelling’s $T^2$ statistic; SPE denotes squared prediction error; $D_r$ and $D_p$ denote SPE and $T^2$ of statistics patterns (SPs) in SPA framework, respectively. For SPA, the selected statistics are mean, variance, skewness, and kurtosis. For DPCA, SPA and RDTCSA, the time lag is set to 2, 1 and 1 respectively. The window lengths are all set as the commonly used 100. For RTCSA, RDTCSA and PMIM, $\ell_2$ norm is used as scalarization. The significance level is set as 5\%.
\end{tablenotes}
\end{threeparttable}
\vspace{-.0in}
\end{table}
}
{
\linespread{1}
\begin{table}[!hbpt]
\small
\caption{The FARs $(\%)$ of different methods for the numerical simulations} \label{1:TableFAR}
\centering
\begin{threeparttable}
\renewcommand{\arraystretch}{1.1}
\renewcommand{\tabcolsep}{2mm}
\begin{tabular}{c||c|c||c|c||c||c||c}\hline
\toprule
    \textbf{No.} &\multicolumn{2}{|c||}{\textbf{DPCA}}
&\multicolumn{2}{|c||}{\textbf{SPA}} &\textbf{RTCSA} &\textbf{RDTCSA} &\textbf{PMIM} \\\cline{2-3}\cline{4-5}
&\footnotesize{$T^{2}$} &\footnotesize{SPE} &\footnotesize{$D_{r}$} &\footnotesize{$D_{p}$}  &  &  &  \\\hline
\midrule
  1 &17.31 &18.28   &0.22  &10.32  &6.22  &3.11  &\textbf{1.78} \\  \hline
  2 &20.20 &19.44   &0  &0  &4.67  &\textbf{1.44}  &5.01  \\  \hline
  3 &18.28 &15.53   &0  &9.54  &4.88  &3.65  &\textbf{2.77}   \\  \hline
  4 &19.44 &17.92   &0  &15.54  &11.88 &15.53  &\textbf{2.77}   \\  \hline
  \textbf{Aver.}    &18.81  &17.79 &0.055 &8.85  &6.91  &5.93  &\textbf{3.08}  \\  \hline
\bottomrule
\end{tabular}
\end{threeparttable}
\vspace{-.0in}
\end{table}
}

According to Table~\ref{1:TableFDR}, PMIM can effectively detect different types of faults and has the highest detection rate. Our advantage becomes more obvious for fault Type III and fault Type V, namely the additive process fault and dynamic changes. Moreover, as demonstrated in Table~\ref{1:TableFAR}, for each test process, PMIM achieves smaller FAR values at the early stage of the normal phase. Although SPA achieves nearly zero FAR values, its FDR values is too small, which indicates that SPA is hard to identify faults here.
This is not hard to understand. Note that SPA uses a time lag of $1$. In this sense, any two adjacent windows of data only differ in $1$ sample. The highly overlapped windows will lead to highly correlated SPs, which severely deteriorate the capability of SPA~\cite{Wang_2010}.

\subsection{TEP Experiment}

As a public benchmark of chemical industrial process, Tennessee Eastman process (TEP) created by the Eastman Chemical Company has been widely used for multivariable process control problems \cite{Downs_1993,Ricker_1995} (see Appendix B on the introduction of TEP process). In this application, we use the simulation data generated by the closed-loop Simulink models developed by Braatz~\cite{Ricker_1995, Chiang_2001,Russell_2012} to evaluate the effectiveness of our proposed PMIM. We use $22$ continuous process measurements (sampled with a sampling interval of 3 minutes) and $11$ manipulated variables (generated at time delay that varys from 6 to 15 minutes) for monitoring, which constitutes $33$ dimensional of input data. To obtain a reliable significance level, we generate $200$ hours of training data ($4,000$ samples in total) and $100$ hours of testing data ($2,000$ samples in total). In each test data, a fault occurs exactly after $20$ hours from the beginning.

\begin{figure*}[!t]
\setlength{\abovecaptionskip}{0.cm}
\setlength{\belowcaptionskip}{-0.0cm}
\centering
\subfigure[Normal (t=$500$)] {\includegraphics[width=.51\textwidth]{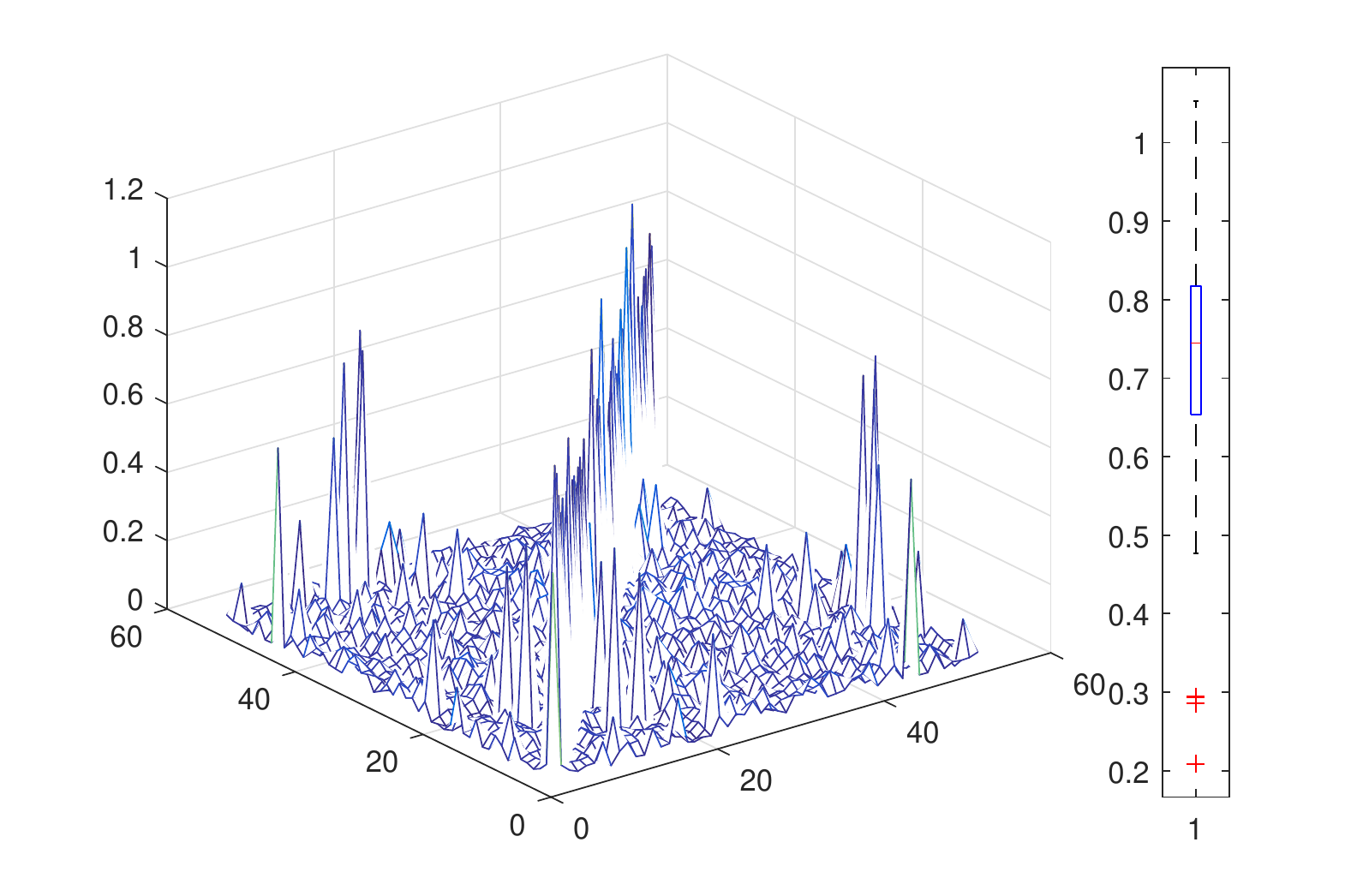}}\hspace{-4mm}
\subfigure[Normal (t=$1,500$)] {\includegraphics[width=.51\textwidth]{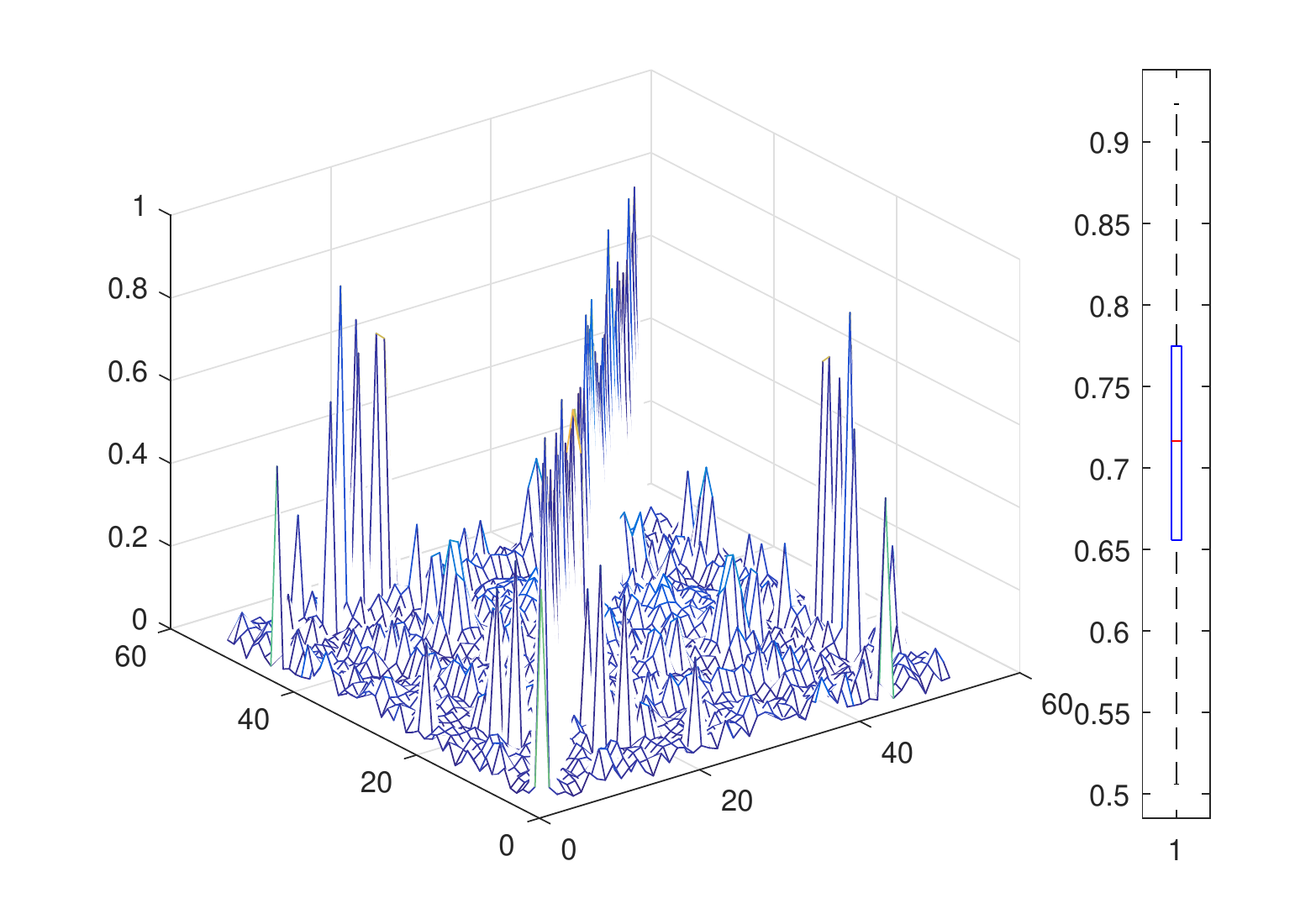}}\hspace{-2mm}
\subfigure[Fault 1 (t=$1,500$)] {\includegraphics[width=.51\textwidth]{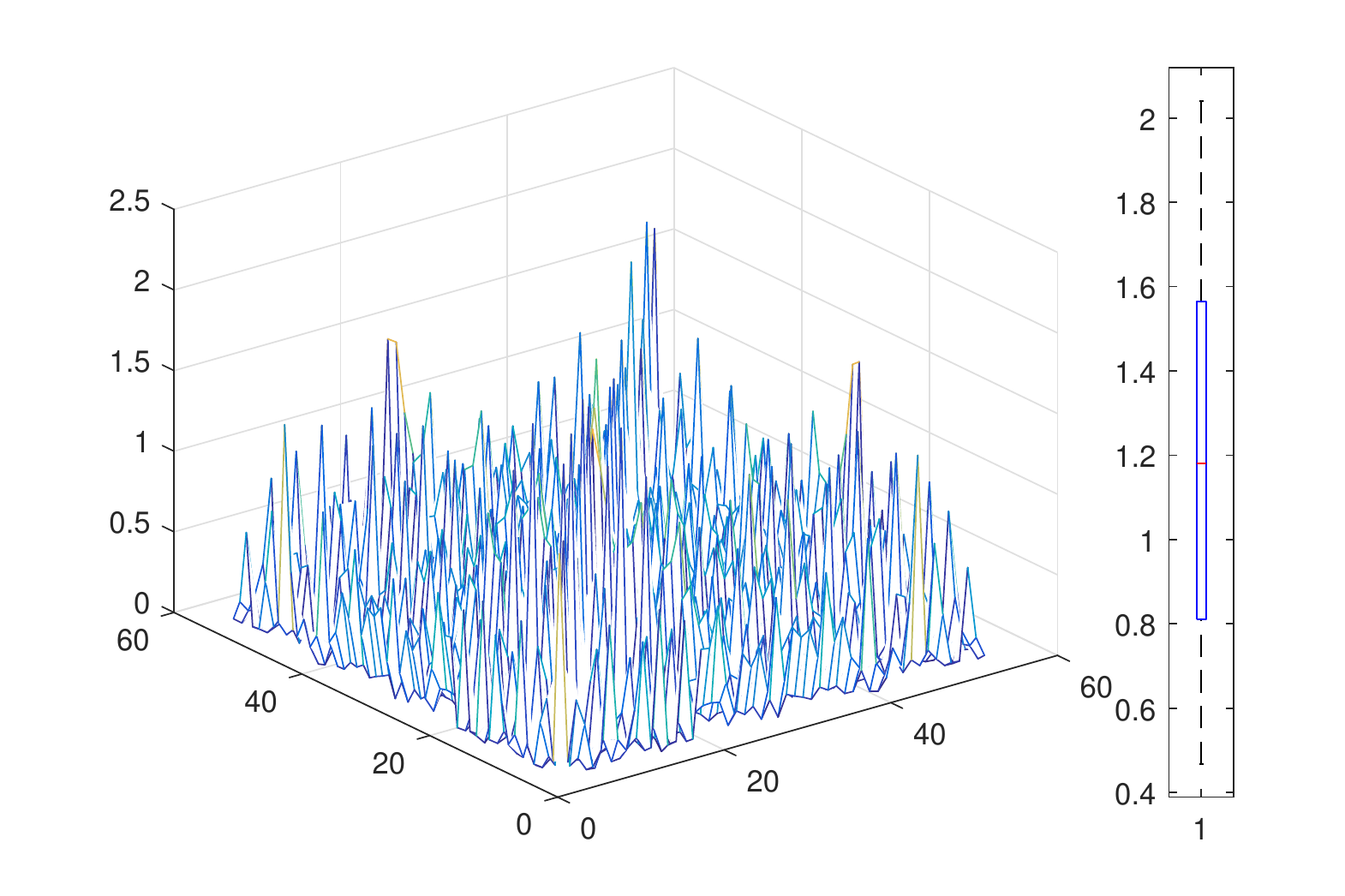}}\hspace{-4mm}
\subfigure[Fault 14 (t=$1,500$)] {\includegraphics[width=.51\textwidth]{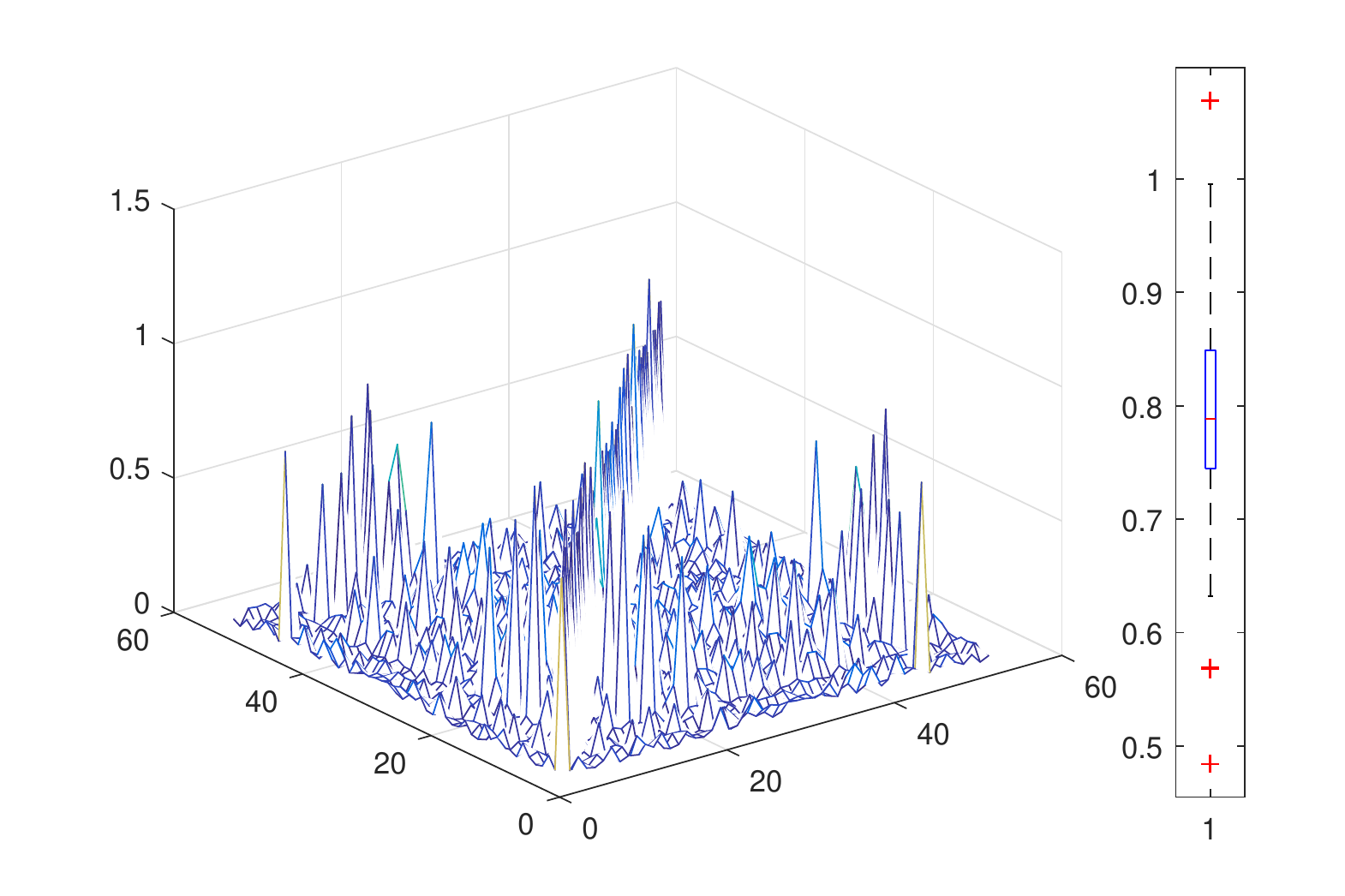}}\hspace{-2mm}
\caption{The MI matrix of TEP under normal and fault states: (a) the MI matrix of normal state at $500$-th sampling instant; (b) the MI matrix of normal state at $1,500$-th sampling instant; (c) the MI matrix of fault 1 at $1,500$-th sampling instant; and (d) the MI matrix of fault 14 at $1,500$-th sampling instant.}
\label{fig:V5}
\end{figure*}

First, the MI matrix (with the boxplot of its diagonal vector) of normal state, fault 1 (step fault) and fault 14 (sticking fault) are shown in Fig.~\ref{fig:V5}. Obviously, the MI matrix keeps almost the same in different time instants under the normal state. However, the occurrence of a fault will lead to different joint or marginal distributions on each dimensional of input, and thus change the entry values in MI matrix. By comparing the boxplots of normal and fauts states, we can observe the changes of diagonal vector, i.e., changes of entropy. Moreover, different types of faults produce different changes of MI matrix.

%\textcolor{blue}{Obviously, there are more outliers, which are plotted individually using the '+' symbol, in Fig.~\ref{fig:V5} (d) than Fig.~\ref{fig:V5} (c).}

\begin{figure*}[!t]
\setlength{\abovecaptionskip}{0.cm}
\setlength{\belowcaptionskip}{-0.0cm}
\centering
\subfigure[Fault 1] {\includegraphics[width=.49\textwidth]{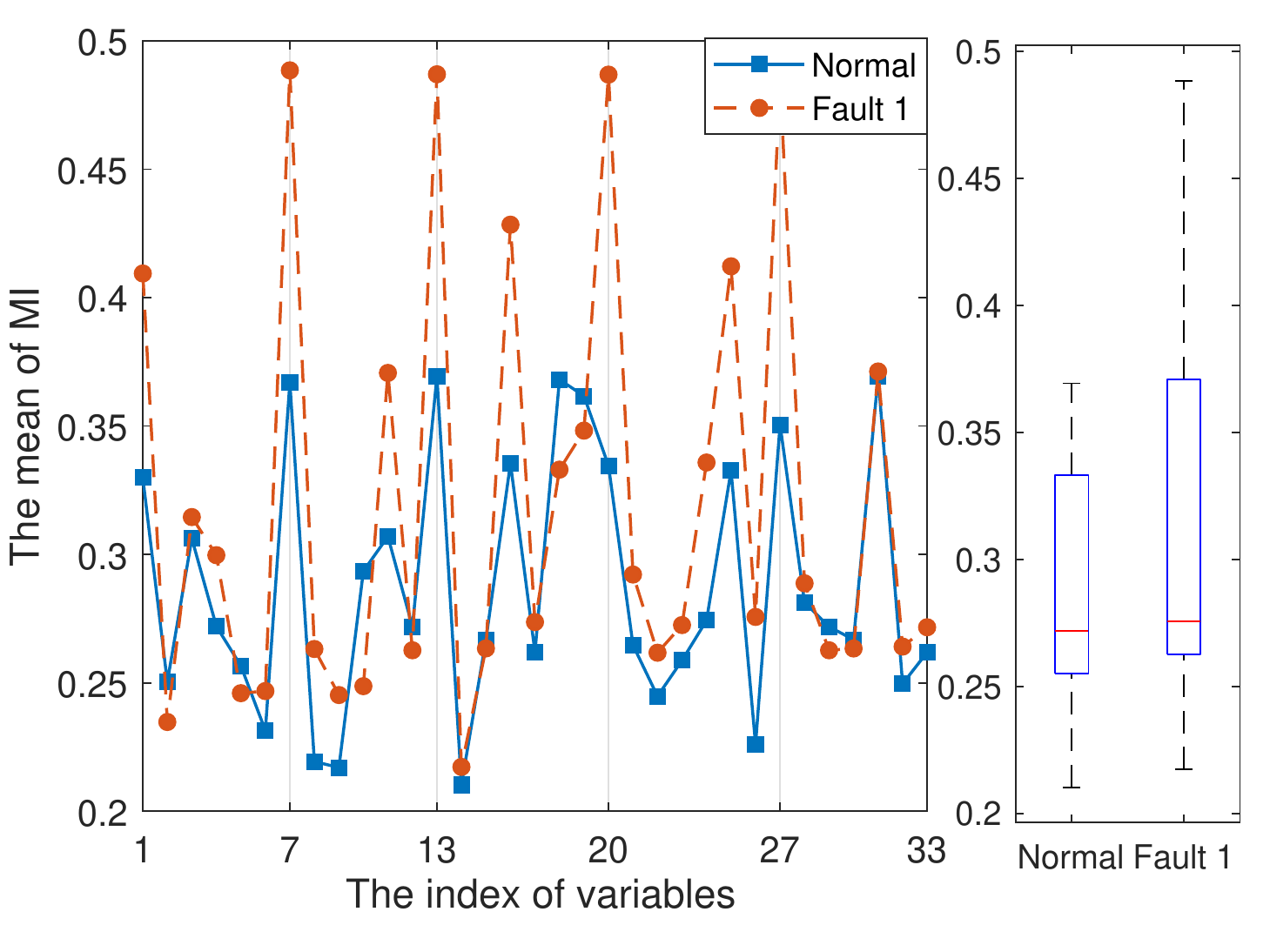}}
\subfigure[Fault 14] {\includegraphics[width=.49\textwidth]{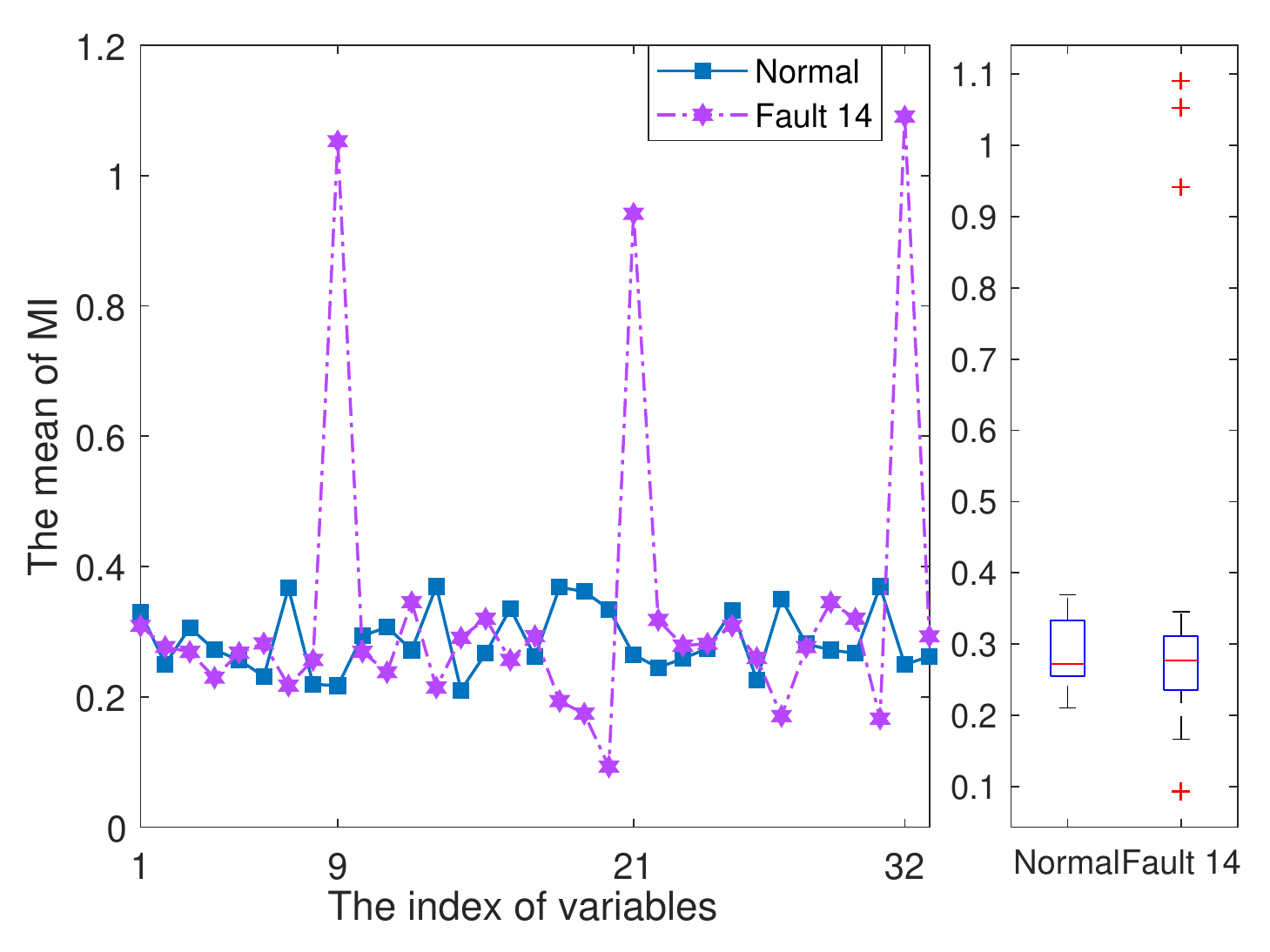}}
\caption{The means of MI matrix of TEP under fault states: (a) fault 1 (step fault); and (b) fault 14 (sticking fault). The left plot is the means of MI along each variable, and the right is their confidence interval.}
\label{fig:V51}
\end{figure*}

The mean of MI values between one variable and all remaining variables\footnote{For the $i$-th variable, we just compute the mean of $I(\mathbf{x}_1,\mathbf{x}_i),\cdots, I(\mathbf{x}_{i-1},\mathbf{x}_i), I(\mathbf{x}_{i+1},\mathbf{x}_i),\cdots, I(\mathbf{x}_m,\mathbf{x}_i)$.} are shown in Fig.~\ref{fig:V51}. As Fig.~\ref{fig:V51}(a) shown, the central box becomes wider and the $75$-th percentiles becomes larger. This indicates that the fault 1 is possibly a step change. In fact, fault 1 indeed induce a step change on stream 4. This feeding changes of reactants A, B and C causes a global impacts on measurements. By contrast, fault 14 induces a sticking change on the reactor cooling water valve, and the most relevant variables are in dimensions $9$, $21$ and $32$~\cite{Chiang_2001}. From Fig.~\ref{fig:V51}(b), there are indeed three outliers which are plotted individually using the $``+"$ symbol, corresponding to the $9$-th, $21$-th and $32$-th dimensional variables.
In other words, the changes on the dimensions $9$, $21$ and $32$ are exactly the driving force that lead to the changes in MI matrix (and hence its eigenspectrum). In this sense, our PMIM also provides insights on the exact root variables that cause the fault, i.e., our fault detection using PMIM is interpretable. One should also note that, an interpretable results also benefit problems related to fault isolation~\cite{ChenZ2016} and restoration~\cite{LiG2011}.

% \textcolor{blue}{, containing the means of MI along each variable (left), and  their confidence interval (right).}

Next, we use the empirical method to determine the confidence limits of different MSPM methods under the same confidence level. Without loss of generality, the window lengths of all competing methods are set to $100$, and all the statics mentioned in Section 3 are used here. The average FDR and FAR values of different MSPM methods on TEP are summarized in Table~\ref{1:Tableeigen2} and Table~\ref{1:Tableeigen1}, respectively.

\begin{figure}[!hbpt]
\centering
\includegraphics[width=0.5\textwidth]{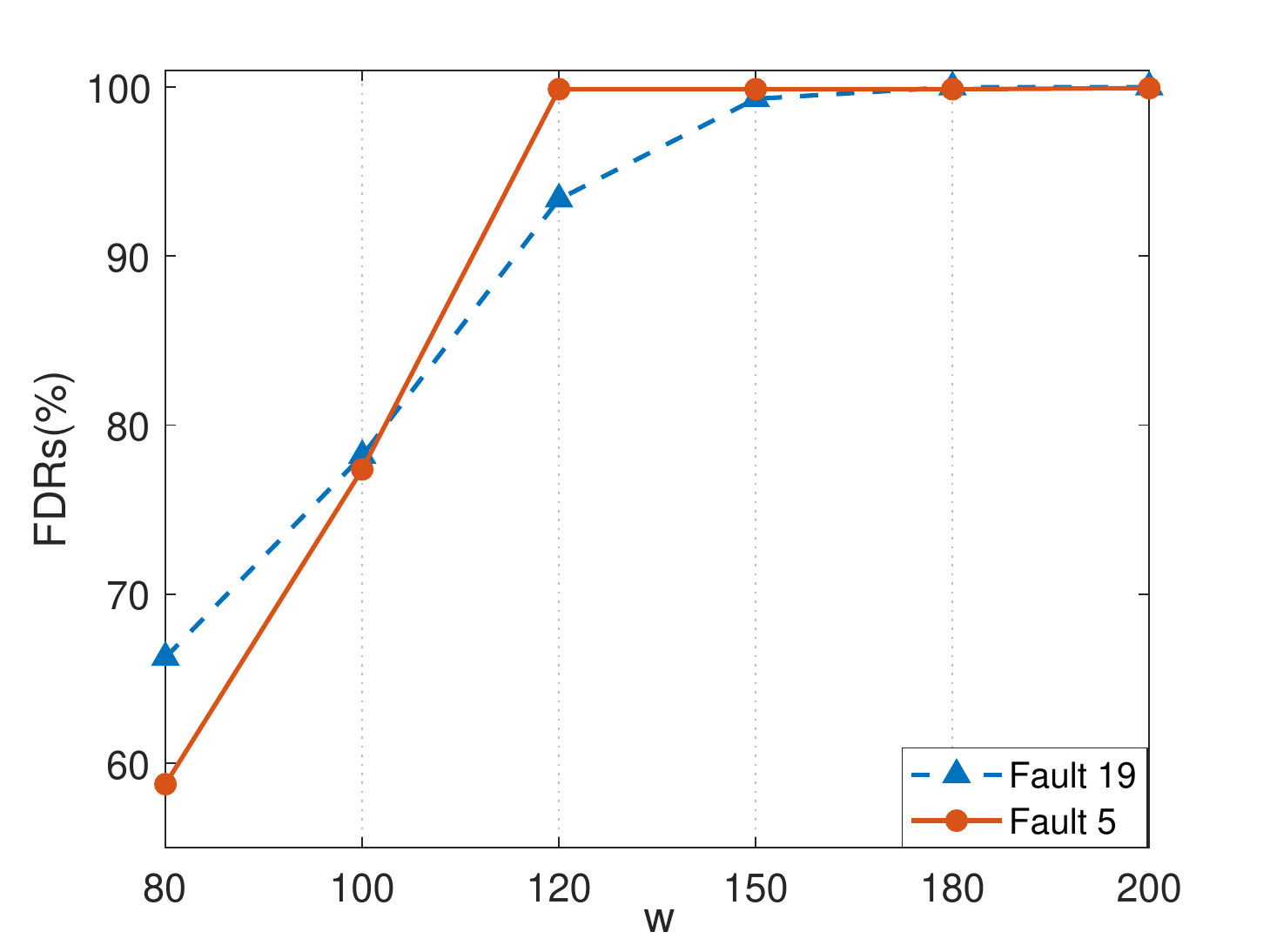}\\
\caption{Detection performances in terms of FDR of different $w$ for fault 5 and 19 in TEP. $w\in\{80,~100,~120,$ $~150,180,200\}$. Fault 5 is marked by red, fault 19 is marked by blue.}
\label{fig:Diffw}
\end{figure}

It can be observed from Table~\ref{1:Tableeigen2} that the FDR of RTCSA, RDTCSA, and PMIM are consistently higher than other methods and remain stable across different types of faults. Moreover, our PMIM always outperforms RTCSA, owing to the superiority of MI over covariance matrix in capturing the intrinsic interactions (either linear or non-linear) between pairwise variables. PMIM detects most of faults. Although our method has relatively lower FDR on step fault 5 and unknown fault 19 with $w=100$, its detection performance in both faults can be significantly improved with larger window size $w$ (see Fig.~\ref{fig:Diffw}.) Detection performances in terms of FDR of different $w$ for fault 5 and 19 are shown in Fig.~\ref{fig:Diffw}. $w=150$ is better to achieve higher FDRs here.

{
\linespread{1}
\begin{table}[!hbpt]
\small
\centering
\begin{threeparttable}
\caption{The FDRs $(\%)$ of different MSPM methods for TEP} \label{1:Tableeigen2}
\renewcommand{\arraystretch}{1.1}
\renewcommand{\tabcolsep}{2.5mm}
\begin{tabular}{l||c|c||c|c||c||c||c}\hline
\toprule
\textbf{No.} &\multicolumn{2}{c||}{\textbf{DPCA}}
&\multicolumn{2}{c||}{\textbf{SPA}} &\textbf{RTCSA} &\textbf{RDTCSA} &\textbf{PMIM} \\ \cline{2-3} \cline{4-5}
\footnotesize{\text{(fault type)}} &\footnotesize{$T^{2}$} &\footnotesize{SPE} &\footnotesize{$D_{r}$} &\footnotesize{$D_{p}$} &  &  & \\       \hline
\midrule
  1 Step &99.91 &\textbf{99.94} &99.88  &99.81  &99.62 &99.56 &99.69  \\  \hline
  2 Step &\textbf{99.19} &98.88 &99.12  &99.12  &98.50 &98.69 &98.31  \\  \hline
  %3 & Step &4.51  &3.63  &6.75  &0.75  &37.0  &64.88 &   \\  \hline
  4 Step &11.63  &\textbf{100} &16.50 &\textbf{100}  &98.38 &99.44 &99.56 \\  \hline
  5 Step &14.94 &28.56 &19.50  &87.81  &\textbf{99.88} &97.25 &77.38   \\  \hline
  6 Step  &99.50 &\textbf{100} &13.63 &13.63  &\textbf{100} &99.94 &\textbf{100}   \\  \hline
  7 Step &\textbf{100}   &\textbf{100} &44.12 &\textbf{100}  &\textbf{100} &\textbf{100} &\textbf{100}   \\  \hline
  8 Random &98.88 &93.63 &\textbf{99.12} &\textbf{99.12}  &97.88 &97.75 &98.62 \\  \hline
  %9 & Random variation &3.50  &2.63 &4.64 &0.62 &32.38 &67.88 &\textbf{70.50}   \\  \hline
  10 Random &21.69 &51.62 &59.56 &88.12  &\textbf{96.63} &37.38 &96.06   \\  \hline
  11 Random &36.88 &95.44 &99.69 &\textbf{100}  &96.25 &92.94 &99.0   \\  \hline
  12 Random &99.38 &97.31 &99.31 &99.31  &99.38 &99.50 &\textbf{100}   \\  \hline
  13 Slow drift &98.56 &92.31 &98.31 &\textbf{100}  &97.88 &98.0 &98.25   \\  \hline
  14 Sticking &99.88 &\textbf{99.94} &\textbf{99.94} &\textbf{99.94}  &99.88 &99.88 &99.88   \\  \hline
  %15 & Sticking &3.13  &4.13 &11.12 &3.62 &37.13 &42.63 &\textbf{86.63}   \\  \hline
  16 Unknown &15.37 &52.38 &63.56 &91.81  &\textbf{99.75} &79.31 &99.50   \\  \hline
  17 Unknown &87.19 &98.31 &98.0 &\textbf{99.31}  &97.81 &97.75 &97.88   \\  \hline
  18 Unknown &94.56 &\textbf{95.75} &93.81 &95.56  &93.75 &93.69 &94.69   \\  \hline
  19 Unknown &48.25 &49.75 &29.38 &99.62  &\textbf{100} &97.19 &78.19   \\  \hline
  20 Unknown &47.38 &61.31 &96.19 &\textbf{96.75}  &96.69 &95.81 &96.31  \\  \hline
  %21 &46.06 &37.55 &57.5 &35 &70.75 &\textbf{95.50} &93.38   \\  \hline
%\textbf{Aver.} &60.07 &61.71 &80.67 &57.85 &85.01 &92.01 &\textbf{93.28}   \\  \hline
\bottomrule
\end{tabular}
\begin{tablenotes}
\footnotesize
\item The window lengths are all set as 100. The selected statistics are mean, variance, skewness, and kurtosis. For RTCSA, RDTCSA and PMIM, $\ell_{\infty}$ norm is used as scalarization. For DPCA and RDTCSA, the time lag is set to 2 and 1 respectively, recommended by authors~\cite{Shang_2017,Shang_2018}. The significance level is set as 2\%.
\end{tablenotes}
\end{threeparttable}
\vspace{-.0in}
\end{table}
}
{
\linespread{1}
\begin{table}[!hbpt]
\small
\caption{The average FARs $(\%)$ of different MSPM methods for TEP} \label{1:Tableeigen1}
\centering
\renewcommand{\arraystretch}{1.1}
\renewcommand{\tabcolsep}{2mm}
\begin{tabular}{c||c|c||c|c||c||c||c}\hline
\toprule
    \textbf{FAR}&  \multicolumn{2}{|c||}{\textbf{DPCA}}
& \multicolumn{2}{|c||}{\textbf{SPA}} &\textbf{RTCSA} &\textbf{RDTCSA} &\textbf{PMIM} \\ \cline{2-3} \cline{4-5}
\footnotesize{$(\%)$} &\footnotesize{$T^{2}$} &\footnotesize{SPE} &\footnotesize{$D_{r}$} &\footnotesize{$D_{p}$} &  &  &  \\  \hline
\midrule
 Normal &2.05 & 3.95 & 4.73 &5.96 &2.89 &3.63 &\textbf{1.18} \\  \hline
\bottomrule
\end{tabular}
\vspace{-.0in}
\end{table}
}

\begin{figure}[hbpt]
\centering
\includegraphics[width=0.5\textwidth]{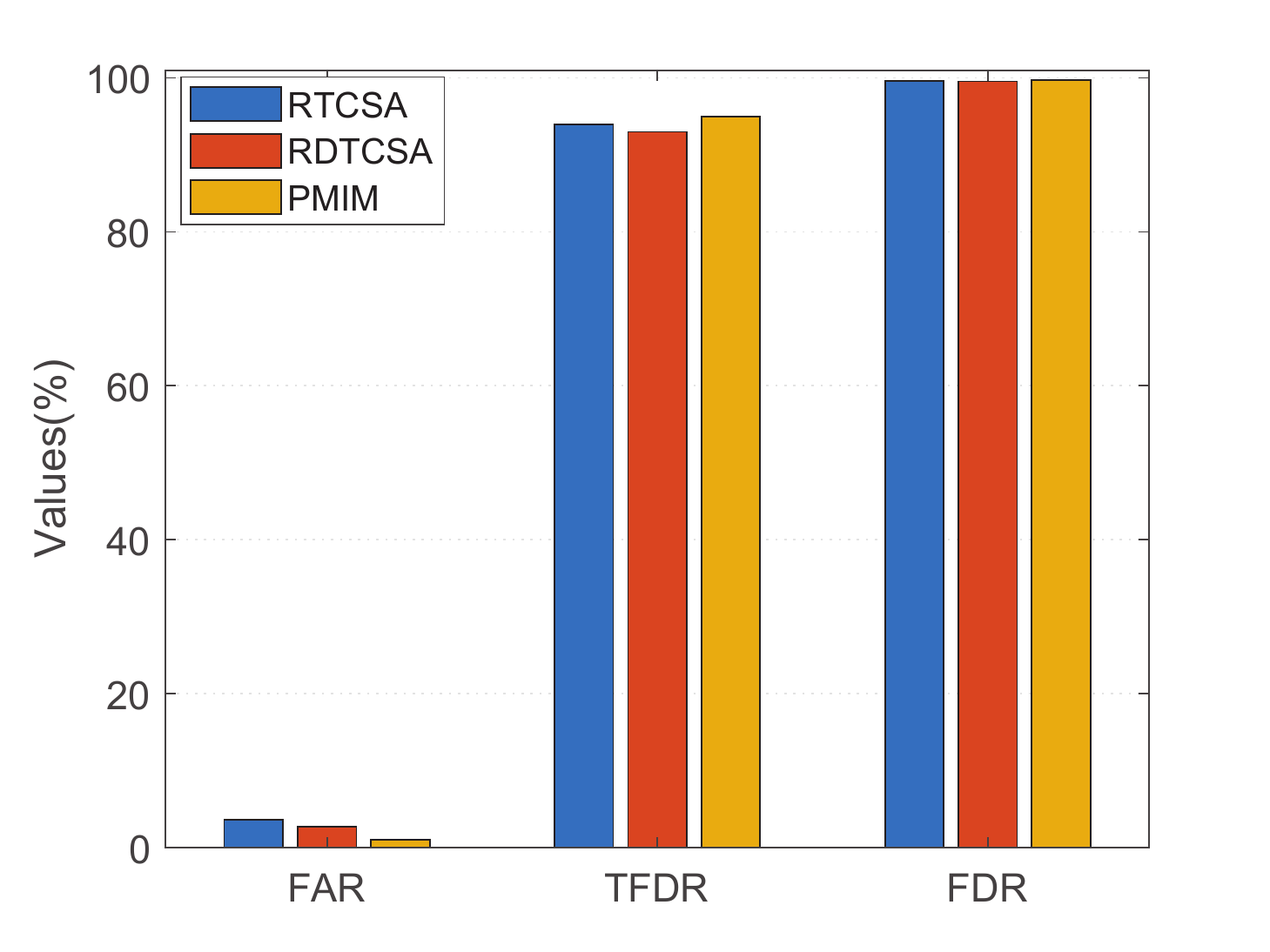}\\
\caption{Detection performances of TCSA methods for fault 1 in TEP. TFDR refers to the FDR value in transition phase. The higher TFDR, the better performance of the used methodology. The methods of RTCSA, RDTCSA and PMIM are marked by blue, red and yellow respectively.}
\label{fig:DTE}
\end{figure}

From Table~\ref{1:Tableeigen1} all the methods achieve favorable FAR, approaching to the theoretical minimum value, i.e., the used significance level. Moreover, our FAR is lower than RTCSA and RDTCSA. This result confirms the superiority of MI in capturing the intrinsic interactions. On the other hand, the detection delay is inevitable owing to the use of sliding windows, a common drawback of the window-based MSPM methods. Take fault 1 for instance, detection performances in terms of FAR, FDR and TFDR (we define the  FDR value in the transition phase\footnote{Transitional phase can be regarded as a connection process between its two neighboring stable phases, in which the window contains both normal and abnormal samples.} as TFDR, the higher the better), of RTCSA, RDTCSA and PMIM are illustrated in Fig.~\ref{fig:DTE}. Our proposed PMIM has the lowest FAR and highest TFDR, which indicates that PMIM is more sensitive to fault 1 than RTCSA and RDTCSA. The detection delay of the proposed method is only 4 samples, which is acceptable in window-based approaches.

\begin{figure}[hbpt]
\centering
\includegraphics[width=1\textwidth]{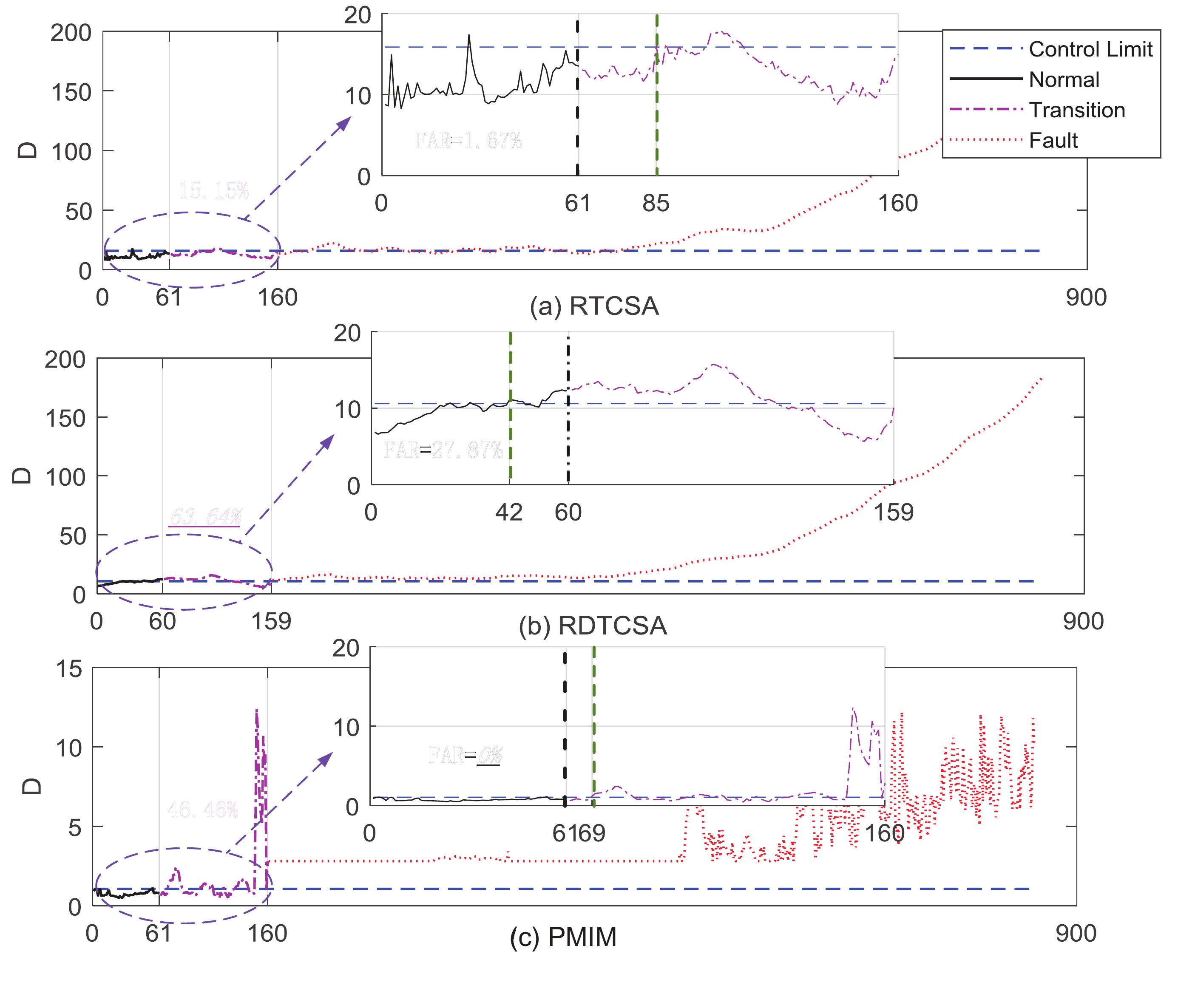}\\
\caption{Detection performances of TCSA methods for fault 21 in TEP. The occurrence of fault corresponded to the $61$-$th$ (RTCSA, PMIM) / $60$-$th$ (RDTCSA) measurements, marked by black line. The FDR values in transition phase are marked by pink. The green line indicates the first sample that detected as a fault instant.}
\label{fig:F21}
\end{figure}

To describe the effectiveness of our proposed PMIM by a more general data, we use the benchmark data of base model that can be downloaded from: \url{http://web.mit.edu/braatzgroup/links.html}. $960$ samples are used as test data. The fault is induced after $8$ hours, which corresponds to the $161$-th samples. Because the length of sliding window is $100$, the fault occurs at the time index $61$ (for RTCSA and PMIM) and $60$ (for RDTCSA). Take fault $21$ as an example, the detection performances of RTCSA, RDTCSA and our PMIM are shown in Fig.~\ref{fig:F21}. The FARs of three competing methods are $1.67\%$ (for RTCSA), $27.87\%$ (for RDTCSA) and $0$ (for PMIM). Obviously, our method has the lowest FAR in this example. RTCSA detects a fault at the $85$-th sample, which suggests a detection delay of $24$ samples. By contrast, our PMIM detects a fault at time index $69$, with a detection delay of only $8$ samples. RDTCSA fails in this example, because it alarms a fault at time index $42$ ($18$ samples ahead of the occurrence of fault), which is a false detection.

\section{Conclusion}

This work presents a new information-theoretic method on fault detection. Before our work, most of the information-theoretic fault detection methods just use mutual information (MI) as a dependence measure to select the most informative dimensions to circumvent the curse of dimensionality. Distinct from these efforts, our method does not perform feature selection. Instead, we constructed a MI matrix to quantify all nonlinear dependencies between pairwise dimensions of data.
We introduced the matrix-based R{\'e}nyi's $\alpha$-order mutual information estimator to estimate the MI value in each entry of the MI matrix. The new estimator avoids the density estimation and is well-suited for complex industrial process. By monitoring different orders of statistics associated with the transformed components of the MI matrix, we demonstrated that our method is able to quickly detect the distributional change of the underlying process, and to identify the root variables that cause the fault. We compared our method with four state-of-the-art fault detection methods on both synthetic data and the real-world Tennessee Eastman process. Empirical results suggest that our method improves the fault detection rate (FDR) and significantly reduces the false alarm rate (FAR). We also presented a thorough analysis on effects of hyper-parameters (e.g., window length $w$ and kernel width $\sigma$) to the performance of our method and illuminated how they control the trade-off between FAR and FDR.

Finally, one should note that the MI matrix is a powerful tool to analyze and discover pairwise interactions in high dimensions of multivariate time series in signal processing, economics and other scientific disciplines. Unfortunately, most of its properties, characteristics, and practical advantages are still largely unknown. This work is a first step to understand the value of non-parametric dependence measures (especially the MI matrix) in monitoring industrial process. We will continue working along this direction to improve the performance of our method and also theoretically explore its fundamental properties.

\section*{Acknowledgment}

This work was supported by the National Natural Science Foundation of China under Grant 61751304, 61933013, 62003004; and the Henan Provincial Science and Technology Research Foundation of China under Grant 202102210125.

\section*{Appendix A}

\definecolor{codegreen}{rgb}{0,0.6,0}
\definecolor{codegray}{rgb}{0.5,0.5,0.5}
\definecolor{codepurple}{rgb}{0.58,0,0.82}
\definecolor{backcolour}{rgb}{0.95,0.95,0.92}
\lstdefinestyle{mystyle}{
    backgroundcolor=\color{backcolour},
    commentstyle=\color{codegreen},
    keywordstyle=\color{magenta},
    numberstyle=\tiny\color{codegray},
    stringstyle=\color{codepurple},
    basicstyle=\ttfamily\footnotesize,
    breakatwhitespace=false,
    breaklines=true,
    captionpos=b,
    keepspaces=true,
    numbers=left,
    numbersep=5pt,
    showspaces=false,
    showstringspaces=false,
    showtabs=false,
    tabsize=2
}

\lstset{style=mystyle}

For reproducible results, we provide key functions (in MATLTB $2019$a) of the proposed PMIM. Specifically, ``mutual\_information\_estimation.m" estimates the matrix-based R{\'e}nyi's $\alpha$-order mutual information (Eq.~\ref{def_mutual}), in which the ``gaussianMatrix.m" evaluates the kernel induced Gram matrix (Eq.~\ref{K_normal}). ``MI\_matrix.m" obtains a series of mutual information matrix at each time instant $k$. ``MITCSA.m" computes the similarity index (Eq.~\ref{similarity}).

\begin{lstlisting}[language=Octave]
function mutual_information = mutual_information_estimation(variable1,variable2,sigma,alpha)
%  variable 1 is i-th dimensional of the process measurement (i-th variable)
%  variable 2 is j-th dimensional of the process measurement (j-th variable)
%% estimate entropy for variable 1
K_x = real(guassianMatrix(variable1,sigma))/size(variable1,1);
[~, L_x] = eig(K_x);
lambda_x = abs(diag(L_x));
H_x = (1/(1-alpha))*log((sum(lambda_x.^alpha)));    

%% estimate entropy for variable 2
K_y = real(guassianMatrix(variable2,sigma))/size(variable2,1);
[~, L_y] = eig(K_y);
lambda_y = abs(diag(L_y));
H_y = (1/(1-alpha))*log((sum(lambda_y.^alpha)));
    
%% estimate joint entropy H(X,Y)
K_xy = K_x.*K_y.*size(variable1,1);
[~,L_xy] = eig(K_xy);
lambda_xy = abs(diag(L_xy));
H_xy =  (1/(1-alpha))*log( (sum(lambda_xy.^alpha)));
    
%% estimate mutual information I(X;Y)
mutual_information = H_x + H_y - H_xy;

end
\end{lstlisting}

\begin{lstlisting}[language=Octave]
function K = guassianMatrix(X,sigma)
G = X*X';
K = bsxfun(@minus, 2*G, diag(G)');
K = exp((1/(2*sigma^2))*bsxfun(@minus, K, diag(G)));

end
\end{lstlisting}

\begin{lstlisting}[language=Octave]
function MImatrixcell = MI_matrix(data,sigma,alpha,MIsize)
% Input:
%      data is the sample matrix X 
%      MIsize is the length of sliding window
%      alpha is the entropy order
%      sigma is the kernel size
% Output:
%      MImatrixcell is a series of mutual information(MI) matrix over the whole process
[nums nums_vars]=size(data); 
[Data, av, st]=zscore(data);  
for k=1:nums-MIsize+1
    dydata=Data(k:k+MIsize-1,:);
%  MImatrix is the MI matrix at time instant k    
    for i=1:nums_vars
          for j=i:nums_vars
               MImatrix(i,j) = mutual_information_estimation(dydata(:,i),dydata(:,j),sigma,alpha);
               MImatrix(j,i) = MImatrix(i,j);
          end
    end
   MImatrixcell{1,k} = MImatrix;
end

end
\end{lstlisting}

\begin{lstlisting}[language=Octave]
function Di = MITCSA(data,MImatrixcell,MIsize)
% Input:
%      data is the sample matrix X
%      MIdata is the MI matrix of data
%      MIsize is the length w of sliding window
% Output:
%      Di is the similarity index 
for i=1:length(MImatrixcell)
    MImatrix=MImatrixcell{1,i}; 
%  Eigen-decomposition of the mutual information(MI) matrix
    [Vet C]=eig(MImatrix,'vector');
% The MI based transform components(TCs)
    T=data{1,i}*Vet;
% The statistic of TCs
    Mu(i,:) = mean(T);% mean
    V(i,:) = sum((T-Mu(i,:)).^2)/MIsize; % variance
    S1(i,:)= sum((T-Mu(i,:)).^3)/MIsize;
    K1(i,:)= sum((T-Mu(i,:)).^4)/MIsize;
    S(i,:) = S1(i,:)./(V(i,:).^(3/2)); % skewness
    K(i,:) = K1(i,:)./(V(i,:).^2)-3;   % kurtosis
end
Oo = [Mu,V,S,K];    
Mu_mu = mean(Mu);% the reference mean
Oo_mu = mean(Oo);  
Oo_sv = std(Oo,1); 
% The calculation of the similarity index 
for i=1:length(MImatrixcell)
      D1 = Oo(i,:)-Oo_mu;
      D = D1./(Oo_sv);
      Di(1,i) = norm(D,inf); 
end

end
\end{lstlisting}

\section*{Appendix B}

\begin{figure}[hbpt]
\centering
\includegraphics[width=1.1\textwidth]{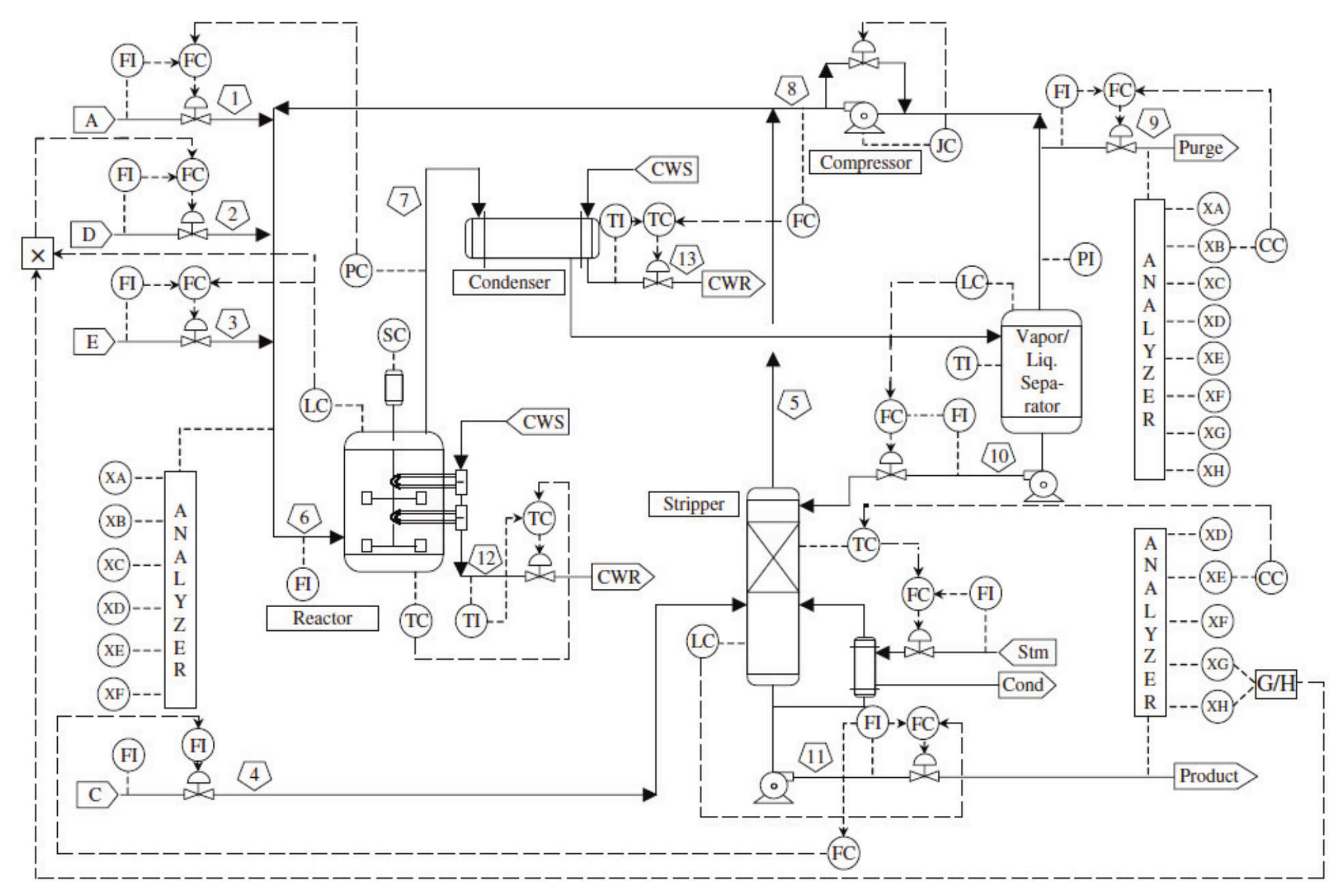}\\
\caption{The schematic of TEP.}
\label{fig:TEP}
\end{figure}

Tennessee Eastman process (TEP) created by the Eastman Chemical Company is designed to provide an actual industrial process for evaluating process control strategies\cite{Downs_1993,Ricker_1995}. It is composed of five major unit operations including a chemical reactor, a product condenser, a recycle compressor, a vapor-liquid separator and a product stripper. Fig.~\ref{fig:TEP} shows its schematic. 21 types of identified faults are listed in Table~\ref{1:Tableeigen3}.  In this work, $33$ different variables ($22$ process measurements and $11$ manipulated measurements) constitute the input of PMIM, as listed in Table~\ref{1:Tableeigen4}. In this sense, the MI matrix in TEP is of size $33\times 33$.

{
\linespread{1}
\begin{table}[!hbpt]
\small
\centering
\begin{threeparttable}
\caption{Descriptions of process faults in TEP} \label{1:Tableeigen3}
\renewcommand{\arraystretch}{1.1}
\renewcommand{\tabcolsep}{2.5mm}
\begin{tabular}{c|c|c}\hline
\toprule
    \textbf{No.} &\textbf{Description} &\textbf{Type} \\ \hline
    \midrule
  1 & A/C feed ratio, B composition constant (Stream 4) & Step \\  \hline
  2 & B composition, A/C ratio constant (Stream 4) & Step \\  \hline
  3 & D feed temperature (Stream 2) & Step  \\  \hline
  4 & Reactor cooling water inlet temperature & Step  \\  \hline
  5 & Condenser cooling water inlet temperature & Step  \\  \hline
  6 & A feed loss (Stream 1) & Step  \\  \hline
  7 & C header pressure loss- reduced availability (Stream 4) & Step  \\  \hline
  8 & A, B, C feed composition (Stream  4)  & Random variation \\  \hline
  9 & D feed temperature (Stream 2) & Random variation  \\  \hline
  10 & C feed temperature (Stream 4) & Random variation  \\  \hline
  11 & Reactor cooling water inlet temperature & Random variation  \\  \hline
  12 & Condenser cooling water inlet temperature & Random variation  \\  \hline
  13 &  Reaction kinetics slow & Slow drift \\  \hline
  14 & Reactor cooling water valve & Sticking  \\  \hline
  15 & Condenser cooling water valve & Sticking  \\  \hline
  16 & Unknown (deviations of heat transfer within stripper (heat exchanger)) & Unknown  \\  \hline
  17 & Unknown (deviations of heat transfer within reactor) & Unknown  \\  \hline
  18 & Unknown (deviations of heat transfer within condenser) & Unknown  \\  \hline
  19 & Unknown & Unknown \\  \hline
  20 & Unknown & Unknown \\  \hline
  21 & The valve for Stream 4 was fixed at the steady state position & Constant position  \\  \hline
\bottomrule
\end{tabular}
\end{threeparttable}
\vspace{-.0in}
\end{table}
}

{
\linespread{1}
\begin{table}[!hbpt]
\small
\begin{threeparttable}
\caption{Monitoring variables in TEP} \label{1:Tableeigen4}
\renewcommand{\arraystretch}{1.1}
\renewcommand{\tabcolsep}{2.5mm}
\begin{tabular}{c c||c c}\hline
\toprule
\textbf{No.} &\textbf{Manipulated measurements} &\textbf{No.} &\textbf{Continuous measurements} \\ \hline
\midrule
1  &D feed flow valve (stream 2) &6 &Reactor feed rate (stream 6)  \\  \hline
2  &E feed flow valve (stream 3) &7 &Reactor pressure  \\  \hline
3  &A feed flow valve (stream 1) &8 &Reactor level  \\  \hline
4  &total feed flow valve (stream 4) &9 &Reactor temperature  \\  \hline
5  &compressor recycle valve &10 &Purge rate (stream 9)  \\  \hline
6  &purge valve (stream 9)  &11 &Product separator temperature   \\   \hline
7  &separator pot liquid flow valve (stream 10)  &12 &Product separator level   \\  \hline
8  &stripper liquid product flow valve (stream 11)  &13 &Product separator pressure   \\  \hline
9  &stripper steam valve   &14 &Product separator underflow  \\  \hline
10 &reactor cooling water flow  &15 &Stripper level   \\  \hline
11 &condenser cooling water flow  &16 &Stripper pressure   \\  \hline
\multicolumn{2}{c||}{\footnotesize{\textbf{Sampling interval: 6 mins}}}  &17 &Stripper underflow \\ \cline{1-2} \hline
\textbf{No.} &\textbf{Continuous measurements}  &18 &Stripper temperature \\  \hline
1 &A feed (stream 1)  &19 &Stripper steam flow   \\  \hline
2 &D feed (stream 2)  &20 &Compressor work    \\  \hline
3 &E feed (stream 3)  &21 &Reactor cooling water outlet temperature    \\  \hline
4 &A and C feed (stream 4)  &22 &Separator cooling water outlet temperature   \\  \hline
5 &Recycle flow (stream 4)  &\multicolumn{2}{c}{\footnotesize{\textbf{Sampling interval: 3 mins}}}    \\ \cline{3-4}   \hline
\bottomrule
\end{tabular}
\end{threeparttable}
\vspace{-.0in}
\end{table}
}

\bibliography{Manuscript}

\end{document}